\documentclass[12pt]{article}
\usepackage{jheppub}
\usepackage{amsfonts,ulem,bbold}   
\usepackage{amssymb,amsmath}

\def\be{\begin{equation}}
\def\ee{\end{equation}}
\def\ba{\begin{eqnarray}}
\def\ea{\end{eqnarray}}

\title{A non-linear extension of the spin-2 partially massless symmetry} 

\author{Sebastian Garcia-Saenz}
\emailAdd{sgsaenz@phys.columbia.edu}
\author{and Rachel A.~Rosen}
\emailAdd{rar2172@columbia.edu}

\affiliation{Physics Department and Institute for Strings, Cosmology, and Astroparticle Physics,\\
  Columbia University, New York, NY 10027, USA}

\abstract{We investigate the possibility of extending the ``partially massless" symmetry of a spin-2 field in de Sitter to nonlinear order.  To do so, we impose a closure condition on the symmetry transformations.  This requirement imposes strong constraints on the form of the nonlinear symmetry while making only minimal assumptions about the form of the nonlinear partially massless action. We find a unique nonlinear extension of the free partially massless symmetry.  However, we show that no consistent Lagrangian that contains at most two derivatives of the fields can realize this symmetry.

}

\begin{document}
\maketitle
\flushbottom
\thispagestyle{empty}
\newpage
%%%%%%%%%%%%%%%%%%%%%%%%%%%%%%%%%%%%%%%%%%%%
%%%%%%%%%%%%%%%%%%%%%%%%%%%%%%%%%%%%%%%%%%%%
\setcounter{page}{1}
\section{Introduction and Summary} \label{sec:intro}

A massive graviton on de Sitter spacetime can acquire an enhanced scalar gauge symmetry for a special choice of the graviton mass relative to the de Sitter curvature. The symmetry removes the helicity-0 mode of the graviton, leaving one fewer physical degrees of freedom. This theory is known as partially massless (PM) gravity \cite{Deser:1983mm,Deser:2001pe,Deser:2001us,Gabadadze:2008uc}.  It enjoys a number of compelling properties such as stability \cite{Higuchi:1986py,Deser:2001wx}, null propagation \cite{Deser:2001xr}, extensions to Einstein backgrounds \cite{Deser:2013bs}, as well as conformal invariance in four dimensions \cite{Deser:2004ji,Deser:2012qg} and a remarkable analogy to electromagnetism \cite{Deser:2006zx,Deser:2013xb,Hinterbichler:2014xga}.

Recently, the question of whether or not a nonlinear theory of PM gravity exists has received much attention \cite{Zinoviev:2006im,deRham:2012kf,Hassan:2012gz,Joung:2012hz,Hassan:2012rq,Joung:2012rv,Deser:2013uy,Hassan:2013pca,Deser:2013gpa,deRham:2013wv,Joung:2014aba,Hassan:2014vja}.  In part, this is due to the recent success in constructing nonlinear ghost-free theories of massive gravity \cite{deRham:2010ik,deRham:2010kj,Hassan:2011hr,deRham:2011qq,deRham:2011rn} (see, \cite{Hinterbichler:2011tt,deRham:2014zqa} for reviews).  Yet despite this progress and some encouraging findings, solid no-go results exist in the literature.  In $D=4$ dimensions (and only $D=4$) there exists a two-derivative cubic Lagrangian with a nonlinear partially massless symmetry \cite{Zinoviev:2006im,deRham:2013wv}.  However, in arbitrary spacetime dimensions, no two-derivative quartic Lagrangian exists \cite{Zinoviev:2006im}.  In particular, among the new nonlinear ghost-free massive gravity theories, no PM theory exists \cite{Deser:2013uy,deRham:2013wv}.

In this paper we will take a somewhat different route from that of previous works to search for a nonlinear PM theory.  We will focusing on the candidate symmetries rather on the candidate Lagrangians of PM gravity.  Our objective is to determine the possible nonlinear extensions of the PM gauge symmetry, without making any {\it a priori} assumptions about the form of the action.  The main tool we will use is a consistency condition on the nonlinear symmetry: we will demand that the symmetry forms an algebra.  In other words, the commutator of two successive transformations should itself be a transformation.  Our analysis largely follows that of Wald \cite{Wald:1986bj} who used this approach to derive the nonlinear symmetries of massless spin-1 and spin-2 fields. 

In particular, our starting point is the partially massless symmetry of the free theory:
\begin{equation} 
\label{introPM}
\delta h_{\mu\nu}^{(0)}=\left(\bar{\nabla}_{\mu}\bar{\nabla}_{\nu}+H^2\,\bar{g}_{\mu\nu}\right)\phi(x) \, .
\end{equation}
We consider a generic nonlinear extension of this symmetry of the form
\begin{equation} 
\label{introPMnl}
\delta h_{\alpha\beta}=B^{\mu\nu}_{\phantom{\mu\nu}\alpha\beta}(\bar{\nabla}_{\mu}\bar{\nabla}_{\nu}+D^{\lambda}_{\phantom{\lambda}\mu\nu}\bar{\nabla}_{\lambda}+C_{\mu\nu}) \phi(x) \, .
\end{equation}
The $B$, $C$, and $D$ tensors are functions of the field $h_{\mu\nu}$ and its derivatives and must reduce to \eqref{introPM} at lowest order in the fields.  The main assumption of this paper is the number of derivatives that appear in the gauge transformation, which we limit to be two.  Thus $B^{\mu\nu}_{\phantom{\mu\nu}\alpha\beta}$ contains no derivatives, $D^{\lambda}_{\phantom{\lambda}\mu\nu}$ contains one, and $C_{\mu\nu}$ contains at most two.  

Consistency requires that the nonlinear partially massless symmetry form an algebra:
\begin{equation} 
\left[\delta_{\phi},\delta_{\psi}\right]h_{\alpha\beta}=\delta_{\chi}h_{\alpha\beta} \, .
\end{equation}
We impose this condition on \eqref{introPMnl} and solve for $B$, $C$, and $D$ order by order.  We find a unique nonlinear extension of the PM symmetry:
\begin{equation}
\delta h_{\alpha\beta}=\left(\bar{\nabla}_{\alpha}\bar{\nabla}_{\beta}+H^2\bar{g}_{\alpha\beta}\right)\phi+\gamma\,\bar{g}_{\alpha\beta}\left[F^{\lambda\mu\nu}F_{\lambda\mu\nu}-\frac{2}{(D-1)}F^{\lambda\mu}_{~~\mu}F_{\lambda\nu}^{~~\nu}\right]\phi \, ,
\end{equation}
where $F_{\lambda\mu\nu}\equiv \bar{\nabla}_{\lambda}h_{\mu\nu}-\bar{\nabla}_{\mu}h_{\lambda\nu}$ and $\gamma$ is a free parameter.

The existence of such a symmetry does not guarantee an invariant Lagrangian.  In section \ref{sec:action} we perform a brute force analysis and show that no consistent Lagrangian that contains at most two derivatives can realize this symmetry.  This is both consistent with previous findings and generalizes them, as we are able to exclude the possibility of two derivative Lagrangians that contain, say, no cubic or quartic terms but are still able to realize a nonlinear PM symmetry at higher order.  We discuss further implications of our results in section \ref{sec:discussion}.

\bigskip

{\it Conventions:} Our choice for the metric signature is $\eta_{\mu\nu}={\mathrm{diag}}(-,+,+,+,\ldots)$. We assume throughout that the number of spacetime dimensions is $D\geq3$. Symmetrizations and antisymmetrizations of indices are defined with unit weight.

\section{Background} \label{sec:pm_gravity}

The dynamics of a free massive spin-2 field $h_{\mu\nu}$ on a $D$-dimensional de Sitter background is described by the Fierz--Pauli action:
\begin{equation}
\begin{split}
S&=\int d^Dx\,\sqrt{-\bar{g}}\bigg[-\frac{1}{2}\bar{\nabla}_{\lambda}h^{\mu\nu}\bar{\nabla}^{\lambda}h_{\mu\nu}+\bar{\nabla}_{\lambda}h^{\mu\nu}\bar{\nabla}_{\mu}h^{\lambda}_{\phantom{\lambda}\nu}-\bar{\nabla}_{\mu}h\bar{\nabla}_{\nu}h^{\mu\nu}+\frac{1}{2}\bar{\nabla}^{\mu}h\bar{\nabla}_{\mu}h\\
&\quad+\frac{\bar{R}}{D}\left(h^{\mu\nu}h_{\mu\nu}-\frac{1}{2}h^2\right)-\frac{m^2}{2}\left(h^{\mu\nu}h_{\mu\nu}-h^2\right)\bigg] \, .
\end{split}
\end{equation}
Here $\bar{g}_{\mu\nu}$, $\bar{\nabla}_{\mu}$ and $\bar{R}$ are the metric, covariant derivative and curvature of the de Sitter background. The helicity-1 components of the field are stable whenever the graviton mass, $m$, satisfies the inequality $m^2>0$ \cite{Deser:2006zx}.  The helicity-0 component is stable provided that $m$ satisfies the Higuchi bound \cite{Higuchi:1986py}:
\begin{equation}
m^2>\frac{(D-2)}{D(D-1)}\bar{R} \, .
\end{equation}
When the Higuchi bound is saturated, the action boasts a scalar gauge symmetry of the form
\begin{equation} 
\label{eq:pm_symmetry}
\delta h_{\mu\nu}=\left(\bar{\nabla}_{\mu}\bar{\nabla}_{\nu}+\frac{m^2}{(D-2)}\,\bar{g}_{\mu\nu}\right)\phi(x) \, .
\end{equation}
This symmetry removes the helicity-0 mode, rendering the number of degrees of freedom to be $D(D-1)/2-2$. This is the free partially massless (PM) theory, whose action is given explicitly by
\begin{equation} \label{eq:pm_quadratic}
\begin{split}
S&=\int d^Dx\,\sqrt{-\bar{g}}\bigg[-\frac{1}{2}\bar{\nabla}_{\lambda}h^{\mu\nu}\bar{\nabla}^{\lambda}h_{\mu\nu}+\bar{\nabla}_{\lambda}h^{\mu\nu}\bar{\nabla}_{\mu}h^{\lambda}_{\phantom{\lambda}\nu}-\bar{\nabla}_{\mu}h\bar{\nabla}_{\nu}h^{\mu\nu}+\frac{1}{2}\bar{\nabla}^{\mu}h\bar{\nabla}_{\mu}h\\
&\quad+\frac{H^2}{2}\left(Dh^{\mu\nu}h_{\mu\nu}-h^2\right)\bigg]\, .
\end{split}
\end{equation}
We have chosen to write $m^2$ and $\bar{R}$ in terms of the Hubble constant $H$ via the relations (including the cosmological constant $\Lambda$ for later use)
\begin{equation}
H^2=\frac{\bar{R}}{D(D-1)}=\frac{m^2}{(D-2)}=\frac{2\Lambda}{(D-1)(D-2)} \, .
\end{equation}
While the PM symmetry exists in AdS as well, having $\bar{R}>0$ ensures the stability of the remaining modes.

The partially massless Lagrangian can be written in terms of an invariant field strength tensor \cite{Skvortsov:2006at,Deser:2006zx}:
\begin{equation} 
\label{eq:tensorF}
F_{\lambda\mu\nu}\equiv \bar{\nabla}_{\lambda}h_{\mu\nu}-\bar{\nabla}_{\mu}h_{\lambda\nu} \, .
\end{equation}
Using the notation $F_{\lambda}\equiv \bar{g}^{\mu\nu}F_{\lambda\mu\nu}$, the action \eqref{eq:pm_symmetry} becomes
\begin{equation}
\label{SPM}
S=-\frac{1}{4}\int d^Dx\,\sqrt{-\bar{g}}\bigg[ F^{\lambda\mu\nu}F_{\lambda\mu\nu}-2F^{\lambda}F_{\lambda}\bigg] \, .
\end{equation}
In fact, any action constructed out of $F_{\lambda\mu\nu}$ and (de Sitter) covariant derivatives of $F_{\lambda\mu\nu}$ will be exactly invariant under the symmetry transformation (\ref{eq:pm_symmetry}).  In particular, we can consider non-linear theories which contain higher powers of $F$.  However, the particular choice of terms in \eqref{SPM} guarantees the theory propagates the right number of degrees of freedom for the partially massless theory and contains no ghosts.

As a free theory, PM gravity \eqref{SPM} is consistent.  However, the recent success in constructing nonlinear theories of massive gravity has prompted the search for a nonlinear PM theory of which the action in (\ref{eq:pm_quadratic}) is but the lowest order term (in a power series in $h_{\mu\nu}$), and of which the transformation in (\ref{eq:pm_symmetry}) is only the $h$-independent part of a nonlinear infinitesimal gauge symmetry.  Finding an extension of this symmetry is the focus of this paper.

\section{Closure condition on the PM gauge symmetry} \label{sec:closure_condition}

\subsection{General argument} \label{subsec:general_argument}
Our goal is to extend the lowest order partially massless gauge symmetry \eqref{eq:pm_symmetry} to a fully non-linear (in $h_{\mu\nu}$), two-derivative scalar gauge symmetry.  Generically, this symmetry can be written as
\begin{equation} 
\label{eq:gauge_transf_full}
\delta_{\phi}h_{\alpha\beta}=B^{\mu\nu}_{\phantom{\mu\nu}\alpha\beta}(\bar{\nabla}_{\mu}\bar{\nabla}_{\nu}\phi+D^{\lambda}_{\phantom{\lambda}\mu\nu}\bar{\nabla}_{\lambda}\phi+C_{\mu\nu}\phi)\, .
\end{equation}
Here $B^{\mu\nu}_{\phantom{\mu\nu}\alpha\beta}=B^{(\mu\nu)}_{\phantom{(\mu\nu)}\alpha\beta}=B^{\mu\nu}_{\phantom{\mu\nu}(\alpha\beta)}$ contains no derivatives of $h_{\mu\nu}$; $D^{\lambda}_{\phantom{\lambda}\mu\nu}=D^{\lambda}_{\phantom{\lambda}(\mu\nu)}$ contains a single derivative of $h_{\mu\nu}$; and $C_{\mu\nu}=C_{(\mu\nu)}$ contains terms linear in $\bar{\nabla}\bar{\nabla}h$, quadratic in $\bar{\nabla}h$, and terms with no derivatives. The assumption that the gauge transformation contains at most two derivatives is perhaps the most important restriction in our argument.   At the zeroth order we have, by assumption,
\begin{equation}
B^{(0)\mu\nu}_{\phantom{(0)\mu\nu}\alpha\beta}=\delta^{\mu}_{(\alpha}\delta^{\nu}_{\beta)},\qquad D^{(0)\lambda}_{\phantom{(0)\lambda}\alpha\beta}=0,\qquad C^{(0)}_{\alpha\beta}=H^2\bar{g}_{\alpha\beta}.
\end{equation}

We adopt as a criterion of consistency that the act of two subsequent symmetry transformations must itself constitute a symmetry transformation.   In other words, for this infinitesimal symmetry to be ``integrable," it must satisfy the closure condition that for {\it any} two gauge parameters $\phi$ and $\psi$ the following equation must hold for some function $\chi$, 
\begin{equation} 
\label{eq:closure_cond_full}
\left[\delta_{\phi},\delta_{\psi}\right]h_{\alpha\beta}=\delta_{\chi}h_{\alpha\beta} \, .
\end{equation}
This condition places very strong constraints on the form of the nonlinear symmetry \eqref{eq:gauge_transf_full}.  Our goal is to solve this equation for the unknown tensors $B^{\mu\nu}_{\phantom{\mu\nu}\alpha\beta}$, $D^{\lambda}_{\phantom{\lambda}\mu\nu}$ and $C_{\mu\nu}$ as series in $h_{\mu\nu}$. To do so write these tensors as well as the unknown gauge function $\chi$ as power series in $h_{\mu\nu}$:
\begin{equation}
B^{\mu\nu}_{\phantom{\mu\nu}\alpha\beta}=\sum_n B^{(n)\mu\nu}_{\phantom{(n)\mu\nu}\alpha\beta},\qquad D^{\lambda}_{\phantom{\lambda}\alpha\beta}=\sum_n D^{(n)\lambda}_{\phantom{(n)\lambda}\alpha\beta},\qquad C_{\alpha\beta}=\sum_n C^{(n)}_{\alpha\beta},\qquad \chi=\sum_n \chi^{(n)},
\end{equation}
and the superscript $(n)$ denotes a term with $n$ powers of the field.  We will  examine eq.\ (\ref{eq:closure_cond_full}) order by order in $h_{\mu\nu}$, to solve for $B^{\mu\nu}_{\phantom{\mu\nu}\alpha\beta}$, $D^{\lambda}_{\phantom{\lambda}\mu\nu}$ and $C_{\mu\nu}$.

Before proceeding, we note that there are several ways in which a non-linear symmetry may be a trivial rewriting of the lowest order symmetry: if it arises from a redefinition of the gauge parameter or if it arises from a redefinition of $h_{\mu\nu}$.  In the first case, to see how this can be dealt with, its helpful to consider the Bianchi identity that follows from the above gauge symmetry:
\begin{equation} \label{eq:bianchi_full}
\bar{\nabla}_{\mu}\bar{\nabla}_{\nu}(B^{\mu\nu}_{\phantom{\mu\nu}\alpha\beta}\mathcal{E}^{\alpha\beta})-\bar{\nabla}_{\lambda}(B^{\mu\nu}_{\phantom{\mu\nu}\alpha\beta}D^{\lambda}_{\phantom{\lambda}\mu\nu}\mathcal{E}^{\alpha\beta})+B^{\mu\nu}_{\phantom{\mu\nu}\alpha\beta}C_{\mu\nu}\mathcal{E}^{\alpha\beta}=0,
\end{equation}
where $\mathcal{E}^{\alpha\beta}$ is the full EOM. We observe that there is a redefinition freedom in the $B$, $D$ and $C$ tensors which leaves this expression invariant. Suppose we let
\ba
\label{transf}
\begin{array}{lcl}
B^{\mu\nu}_{\phantom{\mu\nu}\alpha\beta} &\to& fB^{\mu\nu}_{\phantom{\mu\nu}\alpha\beta} \, , \\
D^{\lambda}_{\phantom{\lambda}\alpha\beta}&\to& D^{\lambda}_{\phantom{\lambda}\alpha\beta}+2f^{-1}\delta^{\lambda}_{(\alpha}\bar{\nabla}_{\beta)}f,\\
C_{\alpha\beta}&\to& C_{\alpha\beta}+f^{-1}\bar{\nabla}_{\alpha}\bar{\nabla}_{\beta}f+f^{-1}D^{\lambda}_{\phantom{\lambda}\alpha\beta}\bar{\nabla}_{\lambda}f.
\end{array}
\ea
where $f$ is an arbitrary function constructed from $h_{\mu\nu}$, with $f|_{h=0}=1$.   Then eq.\ (\ref{eq:bianchi_full}) remains unchanged.  At the level of the gauge transformation (\ref{eq:gauge_transf_full}) this is equivalent to the redefinition of the gauge parameter $\phi\to f\phi$.  In what follows we will use the redefinitions in \eqref{transf} to eliminate spurious nonlinear symmetries.

In addition, redefinitions of the field $h_{\mu\nu}$ can also lead to trivial non-linear symmetries.  Suppose we perform an algebraic field redefinition $h_{\mu\nu}\to \widetilde{h}_{\mu\nu}(h_{\lambda\sigma})$. The EOM then changes as
\begin{equation}
\label{Bredef}
\mathcal{E}^{\alpha\beta}\to \widetilde{\mathcal{E}}^{\alpha\beta}\equiv\frac{\delta S}{\delta \widetilde{h}_{\alpha\beta}}=\mathcal{E}^{\lambda\sigma}\frac{\partial h_{\lambda\sigma}}{\partial \widetilde{h}_{\alpha\beta}} \, .
\end{equation}
Thus, inserting \eqref{Bredef}  into \eqref{eq:bianchi_full}, we see that certain terms in $B^{\mu\nu}_{\phantom{\mu\nu}\alpha\beta}$ can be absorbed by field redefinitions of $h_{\mu\nu}$.  We use this freedom of redefinition to simplify our expressions in the following sections.

We also observe that we can set $B^{\mu\nu}_{\phantom{\mu\nu}\alpha\beta}\to \delta^{\mu}_{(\alpha}\delta^{\nu}_{\beta)}$ provided we can find a field redefinition such that
\begin{equation}
B^{\mu\nu}_{\phantom{\alpha\beta}\lambda\sigma} \frac{\partial \widetilde{h}_{\alpha\beta}}{\partial{h}_{\lambda\sigma}}
=\delta^{\mu}_{(\alpha}\delta^{\nu}_{\beta)} \, .
\end{equation}
The integrability condition for this last equation is
\begin{equation} 
\label{eq:btensor_cond}
B^{\mu\nu}_{\phantom{\mu\nu}\alpha\beta}\frac{\partial B^{\lambda\sigma}_{\phantom{\lambda\sigma}\kappa\rho}}{\partial h_{\alpha\beta}}-B^{\lambda\sigma}_{\phantom{\lambda\sigma}\alpha\beta}\frac{\partial B^{\mu\nu}_{\phantom{\mu\nu}\kappa\rho}}{\partial h_{\alpha\beta}}=0.
\end{equation}
When this equation holds, the nonlinear terms of $B^{\mu\nu}_{\phantom{\mu\nu}\alpha\beta}$ can be completely eliminated via a field redefinition.  We will also take advantage of this condition in the arguments that follow.

Finally, we note that the lowest order partially massless symmetry \eqref{eq:pm_symmetry} closes trivially by itself.  From a symmetry point of view, it is consistent to have non-linear PM actions constructed out of the invariants $F_{\mu\nu\lambda}$ and derivatives of $F_{\mu\nu\lambda}$ that are exactly invariant under \eqref{eq:pm_symmetry}.  (Though these actions are not guaranteed to be ghost-free.)  Here, we look instead for nonlinear extensions of \eqref{eq:pm_symmetry}.

\bigskip

\subsection{Closure condition at order zero} \label{subsec:order0}
We start by using the closure condition to constrain the possible first-order corrections to the partially massless symmetry.  The most general order-one $B$, $D$ and $C$ tensors are given by
\ba
\label{eq:order1tensors}
\begin{array}{lcl}
B^{(1)\mu\nu}_{\phantom{(1)\mu\nu}\alpha\beta}&=& b_1\,\bar{g}^{\mu\nu}h_{\alpha\beta}+b_2\,\bar{g}_{\alpha\beta}h^{\mu\nu}+b_3\,\delta^{(\mu}_{(\alpha}h^{\nu)}_{\beta)}+b_4\,\delta^{\mu}_{(\alpha}\delta^{\nu}_{\beta)}h+b_5\,\bar{g}^{\mu\nu}\bar{g}_{\alpha\beta}h \, ,\\
D^{(1)\lambda}_{\phantom{(1)\lambda}\alpha\beta}&=& d_1\,\bar{\nabla}^{\lambda}h_{\alpha\beta}+d_2\,\bar{\nabla}_{(\alpha}h_{\beta)}^{\phantom{\beta)}\lambda}+d_3\,\bar{g}_{\alpha\beta}\bar{\nabla}_{\rho}h^{\rho\lambda}+d_4\,\bar{g}_{\alpha\beta}\bar{\nabla}^{\lambda}h+d_5\,\delta^{\lambda}_{(\alpha}\bar{\nabla}_{\beta)}h\\
&&+d_6\,\delta^{\lambda}_{(\alpha}\bar{\nabla}^{\rho}h_{\beta)\rho} \, ,\\
C^{(1)}_{\alpha\beta}&=& c_1\,\bar{\nabla}_{\alpha}\bar{\nabla}_{\beta}h+c_2\,\bar{\Box}h_{\alpha\beta}+c_3\,\bar{\nabla}_{\rho}\bar{\nabla}_{(\alpha}h_{\beta)}^{\phantom{\beta)}\rho}+c_4\,\bar{g}_{\alpha\beta}\bar{\nabla}_{\kappa}\bar{\nabla}_{\rho}h^{\kappa\rho}+c_5\,\bar{g}_{\alpha\beta}\bar{\Box}h\\
&&+c_6\,H^2\,h_{\alpha\beta}+c_7\,H^2\,\bar{g}_{\alpha\beta}h \, . \nonumber
\end{array}
\\
\ea
The coefficients of these terms must obey the closure condition at zeroth order in $h_{\mu\nu}$.  The zeroth order part of (\ref{eq:closure_cond_full}) reads
\begin{equation}
\left(\delta^{(0)}_{\phi}\delta^{(1)}_{\psi}-\delta^{(0)}_{\psi}\delta^{(1)}_{\phi}\right)h_{\alpha\beta}=\delta^{(0)}_{\chi^{(0)}}h_{\alpha\beta},
\end{equation}
or more explicitly,
\begin{equation} 
\label{eq:closure_cond_order0}
\begin{split}
&\left(\delta^{(0)}_{\phi}B^{(1)\mu\nu}_{\phantom{(1)\mu\nu}\alpha\beta}\right)\left(\bar{\nabla}_{\mu}\bar{\nabla}_{\nu}\psi+H^2\bar{g}_{\mu\nu}\psi\right)+\left(\delta^{(0)}_{\phi}D^{(1)\lambda}_{\phantom{(1)\lambda}\alpha\beta}\right)\bar{\nabla}_{\lambda}\psi+\left(\delta^{(0)}_{\phi}C^{(1)}_{\alpha\beta}\right)\psi-(\phi\leftrightarrow\psi)\\
&=\left(\bar{\nabla}_{\alpha}\bar{\nabla}_{\beta}+H^2\bar{g}_{\alpha\beta}\right)\chi^{(0)}.
\end{split}
\end{equation}

Some of the coefficients appearing in these expressions can be set to zero by exploiting the redefinition freedoms, as explained above.  First, we have four independent quadratic field redefinitions:
\begin{equation}
h_{\mu\nu}\to \widetilde{h}_{\mu\nu}=h_{\mu\nu}+ m_1\, \bar{g}_{\mu\nu}h^2+m_2 \, \bar{g}_{\mu\nu}h^{\lambda\sigma}h_{\lambda\sigma}+m_3\, hh_{\mu\nu}+m_4\, h^{\lambda}_{\phantom{\lambda}\mu}h_{\nu\lambda}\, ,
\end{equation}
with independent parameters $m_1,m_2,m_3,m_4$.  When combined with $B^{(0)\mu\nu}_{\phantom{(0)\mu\nu}\alpha\beta}=\delta^{\mu}_{(\alpha}\delta^{\nu}_{\beta)}$ in the Bianchi identity (\ref{eq:bianchi_full}), these generate four corresponding trivial tensors for $B^{(1)}$:
\begin{equation}
\frac{\partial \widetilde{h}_{\alpha\beta}}{\partial{h}_{\mu\nu}}=
\delta^{\mu}_{(\alpha}\delta^{\nu}_{\beta)}
+ 2\, m_1 \, \bar{g}^{\mu\nu}\bar{g}_{\alpha\beta}h
+2\, m_2\,\bar{g}_{\alpha\beta}h^{\mu\nu}
+m_3\,\left(\bar{g}^{\mu\nu}h_{\alpha\beta}+\delta^{\mu}_{(\alpha}\delta^{\nu}_{\beta)}h\right)
+2\, m_4\,\delta^{(\mu}_{(\alpha}h^{\nu)}_{\beta)} \, .
\end{equation}
Comparing these with eq.\ (\ref{eq:order1tensors}), we see that we can set $b_2=b_3=b_5=0$, as well as any relation between $b_1$ and $b_4$ other than $b_1=b_4$; we will choose $b_1=-b_4$. Second, another trivial tensor $B^{(1)}$ comes from the choice $f=1+(\mathrm{const.})h$ in the redefinition of the gauge parameter:
\begin{equation}
B^{(0)\mu\nu}_{\phantom{(0)\mu\nu}\alpha\beta}\to fB^{(0)\mu\nu}_{\phantom{(0)\mu\nu}\alpha\beta}=B^{(0)\mu\nu}_{\phantom{(0)\mu\nu}\alpha\beta}+(\mathrm{const.})\delta^{\mu}_{(\alpha}\delta^{\nu}_{\beta)}h.
\end{equation}
Therefore we can choose $b_4=0$, and so $b_1=0$ as well.  In other words, we may use all the redefinition freedom of the field and the gauge parameter to set $B^{(1)\mu\nu}_{\phantom{(1)\mu\nu}\alpha\beta}=0$. 

The remaining parameters $d_1,\,\ldots,d_6$ and $c_1,\ldots,c_7$ are constrained by the closure condition.  We can eliminate the unknown function $\chi^{(0)}$ in (\ref{eq:closure_cond_order0}) by operating on both sides with $\bar{\nabla}_{\sigma}$ and antisymmetrizing over the indices $\sigma$ and $\alpha$ (or, equivalently, over $\sigma$ and $\beta$):
\begin{equation}
\label{eq:closure_cond_order0_curl}
\bar{\nabla}_{[\sigma}\left[ 
\left(\delta^{(0)}_{\phi}D^{(1)\lambda}_{\phantom{(1)\lambda}\alpha]\beta}\right)\bar{\nabla}_{\lambda}\psi+\left(\delta^{(0)}_{\phi}C^{(1)}_{\alpha]\beta}\right)\psi 
-(\phi\leftrightarrow\psi) \right]=0.
\end{equation}
We substitute the expressions in (\ref{eq:order1tensors}) into eq.\ (\ref{eq:closure_cond_order0_curl}) and set the coefficient of every independent term equal to zero. We find six independent solutions for the tensors $D^{(1)}$ and $C^{(1)}$, all of which can be written in terms of the invariant tensor $F$ given in (\ref{eq:tensorF}):
\ba
\label{eq:order1results}
\begin{array}{lcl}
D^{(1)\lambda}_{\phantom{(1)\lambda}\alpha\beta}&=& \alpha_1\,F^{\lambda}_{\phantom{\lambda}(\alpha\beta)}+\alpha_2\,\bar{g}_{\alpha\beta}F^{\lambda}+\alpha_3\,\delta^{\lambda}_{(\alpha}F_{\beta)},\\
C^{(1)}_{\alpha\beta}&=& \beta_1\,\bar{\nabla}_{\rho}F^{\rho}_{\phantom{\rho}(\alpha\beta)}+\beta_2\,\bar{g}_{\alpha\beta}\bar{\nabla}_{\rho}F^{\rho}+\beta_3\,\bar{\nabla}_{(\alpha}F_{\beta)}.
\end{array}
\ea
These are simply all the possible expressions such that $\delta^{(0)}_{\phi}D^{(1)\lambda}_{\phantom{(1)\lambda}\alpha\beta}=0$ and $\delta^{(0)}_{\phi}C^{(1)}_{\alpha\beta}=0$, and so they solve eq.\ (\ref{eq:closure_cond_order0}) with
\begin{equation} \label{eq:chi0_equation}
\delta^{(0)}_{\chi^{(0)}}h_{\alpha\beta}=\left(\bar{\nabla}_{\alpha}\bar{\nabla}_{\beta}+H^2\bar{g}_{\alpha\beta}\right)\chi^{(0)}=0.
\end{equation}
Thus $\chi^{(0)}$ is independent of the gauge parameters $\phi$ and $\psi$.

Some of the terms in (\ref{eq:order1results}) lead to trivial symmetries in the sense that they vanish on the linear EOM. Indeed, the EOM that follows from (\ref{eq:pm_quadratic}) can be written as
\begin{equation}
\begin{split}
\mathcal{E}^{(1)\mu\nu}&=\bar{\Box}h^{\mu\nu}-2\bar{\nabla}_{\lambda}\bar{\nabla}^{(\mu}h^{\nu)\lambda}+\bar{\nabla}^{\mu}\bar{\nabla}^{\nu}h+\bar{g}^{\mu\nu}\left(\bar{\nabla}_{\lambda}\bar{\nabla}_{\sigma}h^{\lambda\sigma}-\bar{\Box}h\right)+H^2\left(Dh^{\mu\nu}-\bar{g}^{\mu\nu}h\right)\\
&=\bar{\nabla}_{\lambda}F^{\lambda(\mu\nu)}+\bar{\nabla}^{(\mu}F^{\nu)}-\bar{g}^{\mu\nu}\bar{\nabla}_{\lambda}F^{\lambda},
\end{split}
\end{equation}
and its trace is simply $\bar{g}_{\mu\nu}\mathcal{E}^{(1)\mu\nu}=-(D-2)\bar{\nabla}_{\lambda}F^{\lambda}$. We can therefore set $\beta_2=0$ and choose either $\beta_1$ or $\beta_3$ to be zero; we will choose to set $\beta_3=0$. In summary, the most general nontrivial tensors $B^{(1)}$, $D^{(1)}$ and $C^{(1)}$ satisfying the zeroth order closure condition are given by
\ba
\label{eq:sym1}
\begin{array}{lcl}
B^{(1)\mu\nu}_{\phantom{(1)\mu\nu}\alpha\beta}&=&0\, ,\\
D^{(1)\lambda}_{\phantom{(1)\lambda}\alpha\beta}&=& \alpha_1\,F^{\lambda}_{\phantom{\lambda}(\alpha\beta)}+\alpha_2\,\bar{g}_{\alpha\beta}F^{\lambda}+\alpha_3\,\delta^{\lambda}_{(\alpha}F_{\beta)} \, ,\\
C^{(1)}_{\alpha\beta}&=& \beta_1\,\bar{\nabla}_{\rho}F^{\rho}_{\phantom{\rho}(\alpha\beta)} \, .
\end{array}
\ea

We can now make contact with results known in the literature.  We see that the $D=4$ non-linear partially massless symmetry found in \cite{Zinoviev:2006im,deRham:2013wv,Joung:2014aba} is a specific case of the coefficients given above, in particular when $\alpha_1 \neq 0$ and all other coefficients are zero.  However, we will see that this symmetry does not survive the imposition of the closure condition at higher-orders.

\bigskip

\subsection{Closure condition at order one} \label{subsec:order1}
We now turn to the quadratic terms in the gauge symmetry $B^{(2)}$, $D^{(2)}$ and $C^{(2)}$.  We consider the most general possible form for the transformation:
\ba
\begin{array}{lcl}
B^{(2)\mu\nu}_{\phantom{(2)\mu\nu}\alpha\beta}&=&u_1\,h^{\mu}_{(\alpha}h^{\nu}_{\beta)}+u_2\,hh^{(\mu}_{(\alpha}\delta^{\nu)}_{\beta)}+u_3\,h^2\delta^{\mu}_{(\alpha}\delta^{\nu}_{\beta)}+u_4\,h^{\lambda\sigma}h_{\lambda\sigma}\delta^{\mu}_{(\alpha}\delta^{\nu}_{\beta)}+u_5\,h^{\mu\nu}h_{\alpha\beta}\\
&&+u_6\,hh^{\mu\nu}\bar{g}_{\alpha\beta}+u_7\,hh_{\alpha\beta}\bar{g}^{\mu\nu} +u_8\,h^{\mu\lambda}h^{\nu}_{\lambda}\bar{g}_{\alpha\beta}+u_9\,h_{\alpha\lambda}h^{\lambda}_{\beta}\bar{g}^{\mu\nu}+u_{10}\,h^2\bar{g}^{\mu\nu}\bar{g}_{\alpha\beta}\\
&&+u_{11}\,h^{\lambda\sigma}h_{\lambda\sigma}\bar{g}^{\mu\nu}\bar{g}_{\alpha\beta}+u_{12}\,h^{\lambda}_{(\alpha}\delta^{(\mu}_{\beta)}h^{\nu)}_{\lambda}\, , \nonumber
\end{array}
\\
\ea
\ba
\begin{array}{lcl}
D^{(2)\lambda}_{\phantom{(2)\lambda}\mu\nu}&=&v_1\,h\bar{\nabla}^{\lambda}h_{\mu\nu}+v_2\,h\bar{\nabla}_{(\mu}h^{\lambda}_{\phantom{\lambda}\nu)}+v_3\,h\bar{\nabla}_{(\mu}h\delta^{\lambda}_{\nu)}+v_4\,h\bar{\nabla}_{\sigma}h^{\sigma}_{\phantom{\sigma}(\mu}\delta^{\lambda}_{\nu)}+v_5\,\bar{g}_{\mu\nu}h\bar{\nabla}^{\lambda}h\\
&&+v_6\,\bar{g}_{\mu\nu}h\bar{\nabla}_{\sigma}h^{\lambda\sigma}+v_7\,h_{\mu\nu}\bar{\nabla}^{\lambda}h+v_8\,h_{\mu\nu}\bar{\nabla}_{\sigma}h^{\lambda\sigma}+v_9\,h^{\rho\sigma}\bar{\nabla}_{\rho}h_{\sigma(\mu}\delta^{\lambda}_{\nu)}\\
&&+v_{10}\,h^{\rho\sigma}\delta^{\lambda}_{(\mu}\bar{\nabla}_{\nu)}h_{\rho\sigma}+v_{11}\,\bar{g}_{\mu\nu}h^{\rho\sigma}\bar{\nabla}^{\lambda}h_{\rho\sigma}+v_{12}\,\bar{g}_{\mu\nu}h^{\rho\sigma}\bar{\nabla}_{\rho}h^{\lambda}_{\phantom{\lambda}\sigma}+v_{13}\,h^{\lambda\sigma}\bar{\nabla}_{\sigma}h_{\mu\nu}\\
&&+v_{14}\,h^{\lambda\sigma}\bar{\nabla}_{(\mu}h_{\nu)\sigma}+v_{15}\,\bar{g}_{\mu\nu}h^{\lambda\sigma}\bar{\nabla}_{\sigma}h+v_{16}\,\bar{g}_{\mu\nu}h^{\lambda\sigma}\bar{\nabla}_{\rho}h^{\rho}_{\phantom{\rho}\sigma}+v_{17}\,h_{\sigma(\mu}\bar{\nabla}^{\sigma}h^{\lambda}_{\phantom{\lambda}\nu)}\\
&&+v_{18}\,h_{\sigma(\mu}\bar{\nabla}_{\nu)}h^{\lambda\sigma}+v_{19}\,h_{\sigma(\mu}\bar{\nabla}^{\lambda}h^{\sigma}_{\phantom{\sigma}\nu)}+v_{20}\,h_{\sigma(\mu}\bar{\nabla}^{\sigma}h\delta^{\lambda}_{\nu)}+v_{21}\,h_{\sigma(\mu}\delta^{\lambda}_{\nu)}\bar{\nabla}_{\rho}h^{\rho\sigma} \\
&&+v_{22}\,h^{\lambda}_{\phantom{\lambda}(\mu}\bar{\nabla}_{\nu)}h+v_{23}\,h^{\lambda}_{\phantom{\lambda}(\mu}\bar{\nabla}^{\sigma}h_{\nu)\sigma}\, , \nonumber
\end{array}
\\
\ea
\ba
\begin{array}{lcl}
C^{(2)}_{\mu\nu}&=&w_1\,h\bar{\nabla}_{\mu}\bar{\nabla}_{\nu}h+w_2\,h\bar{\nabla}_{\sigma}\bar{\nabla}_{(\mu}h^{\sigma}_{\phantom{\sigma}\nu)}+w_3\,h\bar{\Box}h_{\mu\nu}+w_4\,\bar{g}_{\mu\nu}h\bar{\Box}h+w_5\,\bar{g}_{\mu\nu}h\bar{\nabla}_{\lambda}\bar{\nabla}_{\sigma}h^{\lambda\sigma}\\
&&+w_6\,h^{\lambda\sigma}\bar{\nabla}_{(\mu}\bar{\nabla}_{\nu)}h_{\lambda\sigma}+w_7\,h^{\lambda\sigma}\bar{\nabla}_{\lambda}\bar{\nabla}_{(\mu}h_{\nu)\sigma}+w_8\,h^{\lambda\sigma}\bar{\nabla}_{\lambda}\bar{\nabla}_{\sigma}h_{\mu\nu}+w_9\,\bar{g}_{\mu\nu}h^{\lambda\sigma}\bar{\Box}h_{\lambda\sigma}\\
&&+w_{10}\,\bar{g}_{\mu\nu}h^{\lambda\sigma}\bar{\nabla}_{\lambda}\bar{\nabla}_{\sigma}h+w_{11}\,\bar{g}_{\mu\nu}h^{\lambda\sigma}\bar{\nabla}_{\rho}\bar{\nabla}_{\lambda}h^{\rho}_{\phantom{\rho}\sigma}+w_{12}\,h_{\mu\nu}\bar{\Box}h+w_{13}\,h_{\mu\nu}\bar{\nabla}_{\lambda}\bar{\nabla}_{\sigma}h^{\lambda\sigma}\\
&&+w_{14}\,h^{\lambda}_{\phantom{\lambda}(\mu}\bar{\nabla}_{\nu)}\bar{\nabla}_{\lambda}h+w_{15}\,h^{\lambda}_{\phantom{\lambda}(\mu}\bar{\nabla}_{\nu)}\bar{\nabla}^{\sigma}h_{\lambda\sigma}+w_{16}\,h^{\lambda}_{\phantom{\lambda}(\mu}\bar{\Box}h_{\nu)\lambda}+w_{17}\,h_{\lambda(\mu}\bar{\nabla}^{\lambda}\bar{\nabla}^{\sigma}h_{\nu)\sigma}\\
&&+w_{18}\,\bar{\nabla}_{\lambda}h\bar{\nabla}^{\lambda}h_{\mu\nu}+w_{19}\,\bar{\nabla}_{\lambda}h\bar{\nabla}_{(\mu}h^{\lambda}_{\phantom{\lambda}\nu)}+w_{20}\,\bar{g}_{\mu\nu}\bar{\nabla}_{\lambda}h\bar{\nabla}^{\lambda}h+w_{21}\,\bar{g}_{\mu\nu}\bar{\nabla}_{\lambda}h\bar{\nabla}_{\sigma}h^{\lambda\sigma}\\
&&+w_{22}\,\bar{\nabla}_{\sigma}h^{\lambda\sigma}\bar{\nabla}_{\lambda}h_{\mu\nu}+w_{23}\,\bar{\nabla}_{\sigma}h^{\lambda\sigma}\bar{\nabla}_{(\mu}h_{\nu)\lambda}+w_{24}\,\bar{g}_{\mu\nu}\bar{\nabla}_{\sigma}h^{\lambda\sigma}\bar{\nabla}_{\rho}h^{\rho}_{\phantom{\rho}\lambda}\\
&&+w_{25}\,\bar{\nabla}_{(\mu}h\bar{\nabla}_{\nu)}h+w_{26}\,\bar{\nabla}_{(\mu}h\bar{\nabla}^{\lambda}h_{\nu)\lambda}+w_{27}\,\bar{\nabla}_{\lambda}h^{\lambda}_{\phantom{\lambda}(\mu}\bar{\nabla}^{\rho}h_{\nu)\rho}\\
&&+w_{28}\,\bar{g}_{\mu\nu}\bar{\nabla}^{\lambda}h^{\rho\sigma}\bar{\nabla}_{\lambda}h_{\rho\sigma}+w_{29}\,\bar{g}_{\mu\nu}\bar{\nabla}^{\lambda}h^{\rho\sigma}\bar{\nabla}_{\rho}h_{\lambda\sigma}+w_{30}\,\bar{\nabla}_{\lambda}h_{\sigma(\mu}\bar{\nabla}^{\lambda}h^{\sigma}_{\phantom{\sigma}\nu)}\\
&&+w_{31}\,\bar{\nabla}_{\lambda}h_{\sigma(\mu}\bar{\nabla}^{\sigma}h^{\lambda}_{\phantom{\lambda}\nu)}+w_{32}\,\bar{\nabla}_{\lambda}h_{\sigma(\mu}\bar{\nabla}_{\nu)}h^{\lambda\sigma}+w_{33}\,\bar{\nabla}_{(\mu}h^{\lambda\sigma}\bar{\nabla}_{\nu)}h_{\lambda\sigma}\\
&&+w_{34}\,H^2hh_{\mu\nu}+w_{35}\,H^2\bar{g}_{\mu\nu}h^2+w_{36}\,H^2\bar{g}_{\mu\nu}h^{\lambda\sigma}h_{\lambda\sigma}+w_{37}\,H^2h^{\lambda}_{\phantom{\lambda}\mu}h_{\nu\lambda}\,. \nonumber
\end{array}
\\
\ea
These tensors must satisfy the part of the closure condition that is linear in $h_{\mu\nu}$ in eq.\ (\ref{eq:closure_cond_full}):
\begin{equation} 
\label{eq:closure_cond_order1}
\left(\delta^{(0)}_{\psi}\delta^{(2)}_{\phi}-\delta^{(0)}_{\phi}\delta^{(2)}_{\psi}\right)h_{\alpha\beta}+\left(\delta^{(1)}_{\psi}\delta^{(1)}_{\phi}-\delta^{(1)}_{\phi}\delta^{(1)}_{\psi}\right)h_{\alpha\beta}=\delta^{(0)}_{\chi^{(1)}}h_{\alpha\beta}+\delta^{(1)}_{\chi^{(0)}}h_{\alpha\beta}.
\end{equation}
For $\delta^{(1)}_{\phi,\psi}$ we use the symmetry transformation found in the previous section \eqref{eq:sym1}.  This closure condition \eqref{eq:closure_cond_order1} will serve to {\it further} constrain the linear tensors in \eqref{eq:sym1}.

As before, some of the $u$ coefficients can be eliminated via redefinitions in the field and in the gauge parameter. Consider first the choice $f=1+(\mathrm{const.})h^2+(\mathrm{const.})h^{\lambda\sigma}h_{\lambda\sigma}$ in the redefinition $B^{\mu\nu}_{\phantom{\mu\nu}\alpha\beta}\to f B^{\mu\nu}_{\phantom{\mu\nu}\alpha\beta}$. This allows us to choose $u_3=u_4=0$. In addition, there are seven cubic field redefinitions:
\ba
\begin{array}{ll}
h_{\mu\nu}\to \widetilde{h}_{\mu\nu}=&h_{\mu\nu}
+n_1\, \bar{g}_{\mu\nu}h^{\lambda\sigma}h_{\sigma\rho}h^{\rho}_{\phantom{\rho}\lambda}
+n_2\,\bar{g}_{\mu\nu}hh^{\lambda\sigma}h_{\lambda\sigma}
+n_3 \,\bar{g}_{\mu\nu}h^3 \\
&+n_4\,h^2h_{\mu\nu}
+n_5\,hh_{\lambda\mu}h_{\nu}^{\phantom{\nu}\lambda}
+n_6 \,h^{\lambda\sigma}h_{\lambda\mu}h_{\sigma\nu}
+n_7 \,h^{\lambda\sigma}h_{\lambda\sigma}h_{\mu\nu} \, ,
\end{array}
\ea
with independent parameters $n_1, n_2, n_3, n_4, n_5, n_6, n_7$.  These generate seven corresponding trivial tensors for $B^{(2)}$:
\ba
\begin{array}{ll}
\frac{\partial \widetilde{h}_{\alpha\beta}}{\partial{h}_{\mu\nu}}= &\delta^{\mu}_{(\alpha}\delta^{\nu}_{\beta)}
+3\,n_1 \, h^{\mu\lambda}h^{\nu}_{\lambda}\bar{g}_{\alpha\beta}
+n_2 \left(h^{\lambda\sigma}h_{\lambda\sigma}\bar{g}^{\mu\nu}\bar{g}_{\alpha\beta}+2hh^{\mu\nu}\bar{g}_{\alpha\beta} \right)
+3\, n_3 \, h^2\bar{g}^{\mu\nu}\bar{g}_{\alpha\beta}\\
&+n_4 \left( h^2\delta^{\mu}_{(\alpha}\delta^{\nu}_{\beta)}+2hh_{\alpha\beta}\bar{g}^{\mu\nu}\right)
+n_5 \left(h_{\alpha\lambda}h^{\lambda}_{\beta}\bar{g}^{\mu\nu}+2hh^{(\mu}_{(\alpha}\delta^{\nu)}_{\beta)} \right)\\
&+n_6\left(h^{\mu}_{(\alpha}h^{\nu}_{\beta)}+2h^{\lambda}_{(\alpha}\delta^{(\mu}_{\beta)}h^{\nu)}_{\lambda} \right)
+n_7 \left(h^{\lambda\sigma}h_{\lambda\sigma}\delta^{\mu}_{(\alpha}\delta^{\nu}_{\beta)}+2h^{\mu\nu}h_{\alpha\beta}\right) \, .
\end{array}
\ea
We can therefore also set $u_1=u_5=u_7=u_8=u_9=u_{10}=u_{11}=0$, leaving $u_2$, $u_6$ and $u_{12}$ to be determined. 

To simplify \eqref{eq:closure_cond_order1}, we again operate on both sides with $\bar{\nabla}_{\sigma}$ and antisymmetrize over the indices $\sigma$ and $\alpha$.  This eliminates the term involving $\chi^{(1)}$.  Since $\chi^{(0)}$ is independent of $\phi$ and $\psi$, we can consider the trivial case $\phi=\psi=0$.  This yields
\begin{equation}
\label{curlxi0}
\bar{\nabla}_{[\sigma}\left(\delta^{(1)}_{\chi^{(0)}}h_{\alpha]\beta}\right)=0.
\end{equation}
Thus the ``curl" of eq.\ (\ref{eq:closure_cond_order1}) does not involve $\chi^{(0)}$ either.

We now substitute $B^{(2)}$, $D^{(2)}$ and $C^{(2)}$ into the ``curl" of eq.\ (\ref{eq:closure_cond_order1}), along with $D^{(1)}$ and $C^{(1)}$ given in \eqref{eq:order1results}. We set the coefficient of every independent term equal to zero. We find that the closure condition forces the {\it linear} terms $D^{(1)}$ and $C^{(1)}$ to vanish:
\ba
\begin{array}{lcl}
D^{(1)\lambda}_{\phantom{(1)\lambda}\alpha\beta}&=& 0 \, , \\
C^{(1)}_{\alpha\beta}&=&0 \, .
\end{array}
\ea
In particular, the non-linear symmetry found for $D=4$ in \cite{deRham:2013wv} does not survive the higher order closure argument.

In addition, we find that $B^{(2)}$ and $D^{(2)}$ must vanish, while tensor $C^{(2)}$ contains the six independent contractions with two powers of the invariant tensor $F$:
\ba
\label{eq:c2tensor}
\begin{array}{lcl}
B^{(2)\mu\nu}_{\phantom{(2)\mu\nu}\alpha\beta}&=& 0 \, , \\
D^{(2)\lambda}_{\phantom{(1)\lambda}\alpha\beta}&=& 0 \, , \\
C^{(2)}_{\alpha\beta}&=&\gamma_1\,\bar{g}_{\alpha\beta}F^{\lambda}F_{\lambda}+\gamma_2\,F_{\alpha}F_{\beta}+\gamma_3\,F^{\lambda}F_{\lambda(\alpha\beta)} \\&&+\gamma_4\,\bar{g}_{\alpha\beta}F^{\lambda\mu\nu}F_{\lambda\mu\nu}+\gamma_5\,F_{\mu\nu\alpha}F^{\mu\nu}_{\phantom{\mu\nu}\beta}+\gamma_6\,F_{\alpha\mu\nu}F_{\beta}^{\phantom{\beta}\mu\nu}.
\end{array}
\ea
With these results the left-hand side of eq.\ (\ref{eq:closure_cond_order1}) vanishes, and so the function $\chi^{(1)}$ must satisfy
\begin{equation} \label{eq:chi1_equation}
\delta^{(0)}_{\chi^{(1)}}h_{\alpha\beta}=\left(\bar{\nabla}_{\alpha}\bar{\nabla}_{\beta}+H^2\bar{g}_{\alpha\beta}\right)\chi^{(1)}=0.
\end{equation}
We see that $\chi^{(1)}$ is also independent of $\phi$ and $\psi$.

%It will be remembered that we assumed that $D^{(1)}$ and $C^{(1)}$ were not both zero in order to conclude that $\chi^{(0)}=0$. We see now that these tensors must in fact vanish, and so $\chi^{(0)}$ need only satisfy eq.\ (\ref{eq:chi0_equation}); this of course does not affect the above conclusions, since in any case the term involving $\chi^{(0)}$ does not appear in (\ref{eq:closure_cond_order1}).

\bigskip

\subsection{Closure condition at order two} \label{subsec:order2}
The fact that $\delta^{(1)}_{\phi} h_{\mu\nu} = 0$ now greatly simplifies the rest of the analysis.  In fact, we no longer need to use a brute force approach.  To see this, 
consider the part of the closure condition involving two powers of $h_{\mu\nu}$
\begin{equation} 
\label{eq:closure_cond_order2}
\left(\delta^{(0)}_{\psi}\delta^{(3)}_{\phi}-\delta^{(0)}_{\phi}\delta^{(3)}_{\psi}\right)h_{\alpha\beta}=\delta^{(0)}_{\chi^{(2)}}h_{\alpha\beta}+\delta^{(2)}_{\chi^{(0)}}h_{\alpha\beta}.
\end{equation}
The term depending on $\chi^{(0)}$ will be irrelevant for the same reason as in the previous case (see \eqref{curlxi0} and discussion).  Thus we will drop it in what follows.  Writing \eqref{eq:closure_cond_order2} more explicitly we have
\begin{equation} \label{eq:closure_cond_order_2}
\begin{split}
&(\delta^{(0)}_{\phi}B^{(3)\mu\nu}_{\phantom{(3)\mu\nu}\alpha\beta})(\bar{\nabla}_{\mu}\bar{\nabla}_{\nu}\psi+H^2\bar{g}_{\mu\nu}\psi)+\delta^{(0)}_{\phi}D^{(3)\lambda}_{\phantom{(3)\lambda}\alpha\beta}\bar{\nabla}_{\lambda}\psi+\delta^{(0)}_{\phi}C^{(3)}_{\alpha\beta}\psi-(\phi\leftrightarrow\psi)\\
&=\bar{\nabla}_{\alpha}\bar{\nabla}_{\beta}\chi^{(2)}+H^2\bar{g}_{\alpha\beta}\chi^{(2)},
\end{split}
\end{equation}
which must hold for some function $\chi^{(2)}$ given any two functions $\phi$ and $\psi$. Take the particular case in which $\psi$ satisfies $(\bar{\nabla}_{\mu}\bar{\nabla}_{\nu}+H^2\bar{g}_{\mu\nu})\psi=0$. Then eq.\ (\ref{eq:closure_cond_order_2}) reduces to
\begin{equation}
\delta^{(0)}_{\phi}D^{(3)\lambda}_{\phantom{(3)\lambda}\alpha\beta}\bar{\nabla}_{\lambda}\psi+\delta^{(0)}_{\phi}C^{(3)}_{\alpha\beta}\psi=\bar{\nabla}_{\alpha}\bar{\nabla}_{\beta}\chi^{(2)}+H^2\bar{g}_{\alpha\beta}\chi^{(2)}.
\end{equation}
Next, operate on this last equation with $\bar{\nabla}_{\sigma}$ and antisymmetrize over $\sigma$ and $\alpha$. The right-hand side then vanishes, and we are left with
\begin{equation}
\delta^{(0)}_{\phi}\left[\bar{\nabla}_{[\sigma}\left(D^{(3)\lambda}_{\phantom{(3)\lambda}\alpha]\beta}\bar{\nabla}_{\lambda}\psi\right)+\bar{\nabla}_{[\sigma}\left(C^{(3)}_{\alpha]\beta}\psi\right)\right]=0 \, .
\end{equation}
It follows that
\begin{equation}
\quad \delta^{(0)}_{\phi}\left(\bar{\nabla}_{[\sigma}D^{(3)\lambda}_{\phantom{(3)\lambda}\alpha]\beta}+\delta^{\lambda}_{[\sigma}C^{(3)}_{\alpha]\beta}\right)\,\bar{\nabla}_{\lambda}\psi+\delta^{(0)}_{\phi}\left(\bar{\nabla}_{[\sigma}C^{(3)}_{\alpha]\beta}-H^2\bar{g}_{\lambda[\sigma}D^{(3)\lambda}_{\phantom{(3)\lambda}\alpha]\beta}\right)\,\psi=0,
\end{equation}
where we used $\bar{\nabla}_{\lambda}\bar{\nabla}_{\sigma}\psi=-H^2\bar{g}_{\lambda\sigma}\psi$. Now, we have only required $\psi$ to satisfy a second-order equation, which means that we have the freedom to choose $\psi$ and $\bar{\nabla}_{\lambda}\psi$ to be linearly independent. We can then conclude that
\ba
\label{eq:gauge_inv_order2}
\begin{array}{l}
\delta^{(0)}_{\phi}\left(\bar{\nabla}_{[\sigma}D^{(3)\lambda}_{\phantom{(3)\lambda}\alpha]\beta}+\delta^{\lambda}_{[\sigma}C^{(3)}_{\alpha]\beta}\right)=0,\\
\delta^{(0)}_{\phi}\left(\bar{\nabla}_{[\sigma}C^{(3)}_{\alpha]\beta}-H^2\bar{g}_{\lambda[\sigma}D^{(3)\lambda}_{\phantom{(3)\lambda}\alpha]\beta}\right)=0.
\end{array}
\ea
These tell us that the quantities inside the parentheses are gauge invariant under the lowest order transformation: $\delta h_{\mu\nu}=(\bar{\nabla}_{\mu}\bar{\nabla}_{\nu}+H^2\bar{g}_{\mu\nu})\phi$.  In other words, they must be constructed out of the invariant $F_{\mu\nu\lambda}$ and covariant derivatives of this invariant or they must be zero.

Consider the first expression in \eqref{eq:gauge_inv_order2}.  The invariant term contains at most two derivatives and yet is cubic order in fields.  No such term can be constructed out of $F_{\mu\nu\lambda}$ and its derivatives, thus we must have
\be
\bar{\nabla}_{[\sigma}D^{(3)\lambda}_{\phantom{(3)\lambda}\alpha]\beta}+\delta^{\lambda}_{[\sigma}C^{(3)}_{\alpha]\beta} = 0\, .
\ee
Substituting this into the second expression in \eqref{eq:gauge_inv_order2}, we find 
\be
\label{parenth}
\delta^{(0)}_{\phi}\left(\bar{\nabla}_\lambda\bar{\nabla}_{[\sigma}D^{(3)\lambda}_{\phantom{(3)\lambda}\alpha]\beta}+H^2\bar{g}_{\lambda[\sigma}D^{(3)\lambda}_{\phantom{(3)\lambda}\alpha]\beta}\right)=0 \, .
\ee
This term is cubic in derivatives and cubic in fields, thus potentially can be constructed out of $F_{\mu\nu\lambda}$.  However, an explicit calculation shows that this is not the case. Indeed, if we start with the most general expression for $D^{(3)\lambda}_{\phantom{(3)\lambda}\alpha\beta}$ (which contains 61 coefficients to be determined) and impose equation (\ref{parenth}), we find that the only possible result is
\begin{equation}
D^{(3)\lambda}_{\phantom{(3)\lambda}\alpha\beta}=2\delta^{\lambda}_{(\alpha}\bar{\nabla}_{\beta)}\big(\hat{d}_1\,h^{\mu\sigma}h^{\nu}_{\phantom{\nu}\sigma}h_{\mu\nu}+\hat{d}_2\,hh^{\mu\nu}h_{\mu\nu}+\hat{d}_3\,h^3\big),
\end{equation}
for which the term in the parenthesis in \eqref{parenth} vanishes identically. Unsurprisingly, this is precisely the form of $D$ that can be removed by a redefinition \eqref{transf}, with $f =1+f^{(3)}$ and
\begin{equation}
f^{(3)}=\hat{d}_1\,h^{\mu\sigma}h^{\nu}_{\phantom{\nu}\sigma}h_{\mu\nu}+\hat{d}_2\,hh^{\mu\nu}h_{\mu\nu}+\hat{d}_3\,h^3\,.
\end{equation}
We conclude that
\begin{equation}
D^{(3)\lambda}_{\phantom{(3)\lambda}\alpha\beta}=0,\qquad C^{(3)}_{\alpha\beta}=0 \, .
\end{equation}

Finally, let us turn to $B^{(3)\mu\nu}_{\phantom{(3)\mu\nu}\alpha\beta}$.  Having established that  $D^{(3)\lambda}_{\phantom{(3)\lambda}\alpha\beta}=0$ and $C^{(3)}_{\alpha\beta}=0$, the closure condition reads:
\begin{equation}
(\delta^{(0)}_{\phi}B^{(3)\mu\nu}_{\phantom{(3)\mu\nu}\alpha\beta})(\bar{\nabla}_{\mu}\bar{\nabla}_{\nu}\psi+H^2\bar{g}_{\mu\nu}\psi)-(\phi\leftrightarrow\psi)=(\bar{\nabla}_{\alpha}\bar{\nabla}_{\beta}+H^2\bar{g}_{\alpha\beta})\chi^{(2)}\, .
\end{equation}
It follows that
\begin{multline}
\left(\frac{\partial B^{(3)\mu\nu}_{\phantom{(3)\mu\nu}\alpha\beta}}{\partial h_{\kappa\rho}}-\frac{\partial B^{(3)\kappa\rho}_{\phantom{(3)\kappa\rho}\alpha\beta}}{\partial h_{\mu\nu}}\right)(\bar{\nabla}_{\mu}\bar{\nabla}_{\nu}\psi+H^2\bar{g}_{\mu\nu}\psi)(\bar{\nabla}_{\kappa}\bar{\nabla}_{\rho}\phi+H^2\bar{g}_{\kappa\rho}\phi) \\
=(\bar{\nabla}_{\alpha}\bar{\nabla}_{\beta}+H^2\bar{g}_{\alpha\beta})\chi^{(2)}.
\end{multline}
As before, we operate on this equation with $\bar{\nabla}_{\sigma}$ and antisymmetrize over $\sigma$ and $\alpha$, with the result
\ba
\begin{array}{l}
\bar{\nabla}_{[\sigma}\left(\frac{\partial B^{(3)\mu\nu}_{\phantom{(3)\mu\nu}\alpha]\beta}}{\partial h_{\kappa\rho}}-\frac{\partial B^{(3)\kappa\rho}_{\phantom{(3)\kappa\rho}\alpha]\beta}}{\partial h_{\mu\nu}}\right)(\bar{\nabla}_{\mu}\bar{\nabla}_{\nu}\psi+H^2\bar{g}_{\mu\nu}\psi)(\bar{\nabla}_{\kappa}\bar{\nabla}_{\rho}\phi+H^2\bar{g}_{\kappa\rho}\phi)\\
+\left(\frac{\partial B^{(3)\mu\nu}_{\phantom{(3)\mu\nu}\beta[\alpha}}{\partial h_{\kappa\rho}}-\frac{\partial B^{(3)\kappa\rho}_{\phantom{(n+1)\kappa\rho}\beta[\alpha}}{\partial h_{\mu\nu}}\right)\Big(\bar{\nabla}_{\sigma]}(\bar{\nabla}_{\mu}\bar{\nabla}_{\nu}\psi+H^2\bar{g}_{\mu\nu}\psi)\,(\bar{\nabla}_{\kappa}\bar{\nabla}_{\rho}\phi+H^2\bar{g}_{\kappa\rho}\phi)\\
+\bar{\nabla}_{\sigma]}(\bar{\nabla}_{\kappa}\bar{\nabla}_{\rho}\phi+H^2\bar{g}_{\kappa\rho}\phi)\,(\bar{\nabla}_{\mu}\bar{\nabla}_{\nu}\psi+H^2\bar{g}_{\mu\nu}\psi)\Big) 
=0\,.
\end{array}
\ea
By considering special cases for $\phi$ and $\psi$, we see that this expression only holds if
\begin{equation} \label{eq:btensor_cond_2}
\frac{\partial B^{(3)\mu\nu}_{\phantom{(3)\mu\nu}\alpha\beta}}{\partial h_{\kappa\rho}}-\frac{\partial B^{(3)\kappa\rho}_{\phantom{(3)\kappa\rho}\alpha\beta}}{\partial h_{\mu\nu}}=0.
\end{equation}
This is the 2nd order part of (\ref{eq:btensor_cond}), the condition under which the higher-order terms in the tensor $B$ can be eliminated by means of a field redefinition. Eq.\ (\ref{eq:btensor_cond}) then implies that there exists a field redefinition that sets $B^{(3)\mu\nu}_{\phantom{(3)\mu\nu}\alpha\beta}=0$. We have thus shown that the nontrivial $3$rd order part of the PM gauge transformation vanishes:
\begin{equation}
B^{(3)\mu\nu}_{\phantom{(3)\mu\nu}\alpha\beta}=0,\qquad D^{(3)\lambda}_{\phantom{(3)\lambda}\mu\nu}=0,\qquad C^{(3)}_{\mu\nu}=0.
\end{equation}
Consquently,
\begin{equation} \label{eq:chi2_equation}
\delta^{(0)}_{\chi^{(2)}}h_{\alpha\beta}=\left(\bar{\nabla}_{\alpha}\bar{\nabla}_{\beta}+H^2\bar{g}_{\alpha\beta}\right)\chi^{(2)}=0,
\end{equation}
so that $\chi^{(2)}$ can have no dependence on $\phi$ and $\psi$.

\bigskip

\subsection{Closure condition at order three} \label{subsec:order3}

At the next order we have
\begin{equation} \label{eq:closure_cond_order3}
\left(\delta^{(0)}_{\phi}\delta^{(4)}_{\psi}-\delta^{(0)}_{\psi}\delta^{(4)}_{\phi}\right)h_{\alpha\beta}+\left(\delta^{(2)}_{\phi}\delta^{(2)}_{\psi}-\delta^{(2)}_{\psi}\delta^{(2)}_{\phi}\right)h_{\alpha\beta}=\delta^{(0)}_{\chi^{(3)}}h_{\alpha\beta}+\delta^{(2)}_{\chi^{(1)}}h_{\alpha\beta}.
\end{equation}
where $\delta^{(2)}_{\phi}h_{\alpha\beta}=C^{(2)}_{\alpha\beta}\phi$, with $C^{(2)}_{\alpha\beta}$ as given in (\ref{eq:c2tensor}). The last term on the right will again be irrelevant for the same reasons given earlier, since from (\ref{eq:chi1_equation}) we see that $\chi^{(1)}$ is independent of $\phi$ and $\psi$. The second term on the left is given by
\begin{equation}
\begin{split}
\left(\delta^{(2)}_{\psi}\delta^{(2)}_{\phi}-\delta^{(2)}_{\phi}\delta^{(2)}_{\psi}\right)h_{\alpha\beta}&=\delta^{(2)}_{\phi}\left(C^{(2)}_{\alpha\beta}\psi\right)-\delta^{(2)}_{\psi}\left(C^{(2)}_{\alpha\beta}\phi\right)\\
&=\psi\,\frac{\partial C^{(2)}_{\alpha\beta}}{\partial\bar{\nabla}_{\lambda}h_{\mu\nu}}\bar{\nabla}_{\lambda}\left(C^{(2)}_{\mu\nu}\phi\right)-\phi\,\frac{\partial C^{(2)}_{\alpha\beta}}{\partial\bar{\nabla}_{\lambda}h_{\mu\nu}}\bar{\nabla}_{\lambda}\left(C^{(2)}_{\mu\nu}\psi\right)\\
&=\frac{\partial C^{(2)}_{\alpha\beta}}{\partial\bar{\nabla}_{\lambda}h_{\mu\nu}}C^{(2)}_{\mu\nu}\left(\psi\bar{\nabla}_{\lambda}\phi-\phi\bar{\nabla}_{\lambda}\psi\right)
\end{split}
\end{equation}
We operate on this equation with $\bar{\nabla}_{\sigma}$ and antisymmetrize over $\sigma$ and $\alpha$ to get rid of the unknown function $\chi^{(3)}$:
\begin{equation} \label{eq:closure_cond_order3_curl}
\bar{\nabla}_{[\sigma}\left(\delta^{(0)}_{\phi}\delta^{(4)}_{\psi}-\delta^{(0)}_{\psi}\delta^{(4)}_{\phi}\right)h_{\alpha]\beta}+\bar{\nabla}_{[\sigma}\left[\frac{\partial C^{(2)}_{\alpha]\beta}}{\partial\bar{\nabla}_{\lambda}h_{\mu\nu}}C^{(2)}_{\mu\nu}\left(\psi\bar{\nabla}_{\lambda}\phi-\phi\bar{\nabla}_{\lambda}\psi\right)\right]=0.
\end{equation}
This equation must hold for any functions $\phi$ and $\psi$.  Consider the case that $\phi$ and $\psi$ satisfy $(\bar{\nabla}_{\mu}\bar{\nabla}_{\nu}+H^2\bar{g}_{\mu\nu})\phi=0$ and $(\bar{\nabla}_{\mu}\bar{\nabla}_{\nu}+H^2\bar{g}_{\mu\nu})\psi=0$ respectively.  Then we are left with
\begin{equation}
\label{eq:closure_cond_order3_curl2}
\bar{\nabla}_{[\sigma}\left[\frac{\partial C^{(2)}_{\alpha]\beta}}{\partial\bar{\nabla}_{\lambda}h_{\mu\nu}}C^{(2)}_{\mu\nu}\left(\psi\bar{\nabla}_{\lambda}\phi-\phi\bar{\nabla}_{\lambda}\psi\right)\right]=0.
\end{equation}
Again, since we have chosen $\phi$ and $\psi$ to obey second order equations, the functions $\phi$, $\psi$, $\bar{\nabla}\phi$ and $\bar{\nabla}\psi$ are still independent and arbitrary.  Thus the only way \eqref{eq:closure_cond_order3_curl2} can be satisfied is if
\begin{equation}
\frac{\partial C^{(2)}_{\alpha\beta}}{\partial\bar{\nabla}_{\lambda}h_{\mu\nu}}C^{(2)}_{\mu\nu}=0.
\end{equation}
This equation constrains the $\gamma$ coefficients in (\ref{eq:c2tensor}) to be
\begin{equation}
\gamma_1=-\frac{2}{(D-1)}\gamma_4\, , \qquad \gamma_2=0\, ,\qquad\gamma_3=0\, ,\qquad \gamma_5=0\, ,\qquad\gamma_6=0\, .
\end{equation}
The tensor $C^{(2)}_{\alpha\beta}$ must then be given by
\begin{equation}
C^{(2)}_{\alpha\beta}=\gamma_4\,\bar{g}_{\alpha\beta}\left[F^{\lambda\mu\nu}F_{\lambda\mu\nu}- \frac{2}{(D-1)}F^{\lambda}F_{\lambda}\right].
\end{equation}
With this result the second term on the left of (\ref{eq:closure_cond_order3}) vanishes by itself, and so
\begin{equation}
\label{neq3}
\left(\delta^{(0)}_{\phi}\delta^{(4)}_{\psi}-\delta^{(0)}_{\psi}\delta^{(4)}_{\phi}\right)h_{\alpha\beta}=\delta^{(0)}_{\chi^{(3)}}h_{\alpha\beta}.
\end{equation}
In the following section we will give a general proof that an equation of the form \eqref{neq3} implies that
\begin{equation}
B^{(4)\mu\nu}_{\phantom{(4)\mu\nu}\alpha\beta}=0,\qquad D^{(4)\lambda}_{\phantom{(4)\lambda}\alpha\beta}=0,\qquad C^{(4)}_{\alpha\beta}=0,
\end{equation}
and
\begin{equation} \label{eq:chi3_equation}
\delta^{(0)}_{\chi^{(3)}}h_{\alpha\beta}=\left(\bar{\nabla}_{\alpha}\bar{\nabla}_{\beta}+H^2\bar{g}_{\alpha\beta}\right)\chi^{(3)}=0.
\end{equation}
Let us turn to this proof now.

\bigskip

\subsection{Recursive Relation}
Our starting assumption is that
\begin{equation}
\label{as1}
B^{(j)\mu\nu}_{\phantom{(0)\mu\nu}\alpha\beta}=0,\qquad D^{(j)\lambda}_{\phantom{(0)\lambda}\mu\nu}=0,\qquad C^{(j)}_{\mu\nu}=0,
\end{equation}
for all $0<j\leq n$, with $n\geq 3$ and with the exception of $C^{(2)}$.  We assume also that
\begin{equation}
\label{as2}
\delta^{(0)}_{\chi^{(i)}}h_{\alpha\beta}=\left(\bar{\nabla}_{\alpha}\bar{\nabla}_{\beta}+H^2\bar{g}_{\alpha\beta}\right)\chi^{(i)}=0,
\end{equation}
for all $0\leq i<n$. The $n$th order part of the closure condition then reads
\begin{equation}
\left(\delta^{(0)}_{\psi}\delta^{(n+1)}_{\phi}-\delta^{(0)}_{\phi}\delta^{(n+1)}_{\psi}\right)h_{\alpha\beta}=\delta^{(0)}_{\chi^{(n)}}h_{\alpha\beta},
\end{equation}
and we have omitted the possible term $\delta^{(2)}_{\chi^{(n-2)}}h_{\alpha\beta}$, which in any case will be irrelevant by the argument given in subsection \ref{subsec:order2}. Writing this more explicitly we have
\begin{multline} 
\label{eq:closure_cond_order_n}
(\delta^{(0)}_{\phi}B^{(n+1)\mu\nu}_{\phantom{(n+1)\mu\nu}\alpha\beta})(\bar{\nabla}_{\mu}\bar{\nabla}_{\nu}\psi+H^2\bar{g}_{\mu\nu}\psi)\\
+\delta^{(0)}_{\phi}D^{(n+1)\lambda}_{\phantom{(n+1)\lambda}\alpha\beta}\bar{\nabla}_{\lambda}\psi
+\delta^{(0)}_{\phi}C^{(n+1)}_{\alpha\beta}\psi-(\phi\leftrightarrow\psi)
=\bar{\nabla}_{\alpha}\bar{\nabla}_{\beta}\chi^{(n)}+H^2\bar{g}_{\alpha\beta}\chi^{(n)} \, ,
\end{multline}
which must hold for some function $\chi^{(n)}$ given any two functions $\phi$ and $\psi$. Take the particular case in which $\psi$ satisfies $(\bar{\nabla}_{\mu}\bar{\nabla}_{\nu}+H^2\bar{g}_{\mu\nu})\psi=0$. Then eq.\ (\ref{eq:closure_cond_order_n}) reduces to
\begin{equation}
\delta^{(0)}_{\phi}D^{(n+1)\lambda}_{\phantom{(n+1)\lambda}\alpha\beta}\bar{\nabla}_{\lambda}\psi+\delta^{(0)}_{\phi}C^{(n+1)}_{\alpha\beta}\psi=\bar{\nabla}_{\alpha}\bar{\nabla}_{\beta}\chi^{(n)}+H^2\bar{g}_{\alpha\beta}\chi^{(n)}.
\end{equation}
Next, operate on this last equation with $\bar{\nabla}_{\sigma}$ and antisymmetrize over $\sigma$ and $\alpha$. The right-hand side then vanishes, and we are left with
\begin{equation}
\delta^{(0)}_{\phi}\left[\bar{\nabla}_{[\sigma}\left(D^{(n+1)\lambda}_{\phantom{(n+1)\lambda}\alpha]\beta}\bar{\nabla}_{\lambda}\psi\right)+\bar{\nabla}_{[\sigma}\left(C^{(n+1)}_{\alpha]\beta}\psi\right)\right]=0\, .
\end{equation}
It follows that
\begin{equation}
 \delta^{(0)}_{\phi}\left(\bar{\nabla}_{[\sigma}D^{(n+1)\lambda}_{\phantom{(n+1)\lambda}\alpha]\beta}+\delta^{\lambda}_{[\sigma}C^{(n+1)}_{\alpha]\beta}\right)\,\bar{\nabla}_{\lambda}\psi+\delta^{(0)}_{\phi}\left(\bar{\nabla}_{[\sigma}C^{(n+1)}_{\alpha]\beta}-H^2\bar{g}_{\lambda[\sigma}D^{(n+1)\lambda}_{\phantom{(n+1)\lambda}\alpha]\beta}\right)\,\psi=0,
\end{equation}
where we used that $\bar{\nabla}_{\lambda}\bar{\nabla}_{\sigma}\psi=-H^2\bar{g}_{\lambda\sigma}\psi$. Again, we have required that $\psi$ satisfy a second-order equation, which means that we still have the freedom to choose $\psi$ and $\bar{\nabla}_{\lambda}\psi$ to be independent. We can then conclude that
\ba
\label{eq:gauge_inv1}
\begin{array}{l}
\delta^{(0)}_{\phi}\left(\bar{\nabla}_{[\sigma}D^{(n+1)\lambda}_{\phantom{(n+1)\lambda}\alpha]\beta}+\delta^{\lambda}_{[\sigma}C^{(n+1)}_{\alpha]\beta}\right)=0,\\
\delta^{(0)}_{\phi}\left(\bar{\nabla}_{[\sigma}C^{(n+1)}_{\alpha]\beta}-H^2\bar{g}_{\lambda[\sigma}D^{(n+1)\lambda}_{\phantom{(n+1)\lambda}\alpha]\beta}\right)=0.
\end{array}
\ea
In other words, the quantities inside the parentheses are invariant under the lowest order transformation $\delta h_{\mu\nu}=(\bar{\nabla}_{\mu}\bar{\nabla}_{\nu}+H^2\bar{g}_{\mu\nu})\phi$ and thus must be constructed out of the $F$ tensors and their derivatives.  However, these terms have at most three derivatives and yet they are all quartic order or higher in the fields.  Since the $F$ tensors and their derivatives have at least one derivative per field, we know that no such invariants exist.  Thus we must have
\ba
\begin{array}{l}
\bar{\nabla}_{[\sigma}D^{(n+1)\lambda}_{\phantom{(n+1)\lambda}\alpha]\beta}+\delta^{\lambda}_{[\sigma}C^{(n+1)}_{\alpha]\beta}=0,\\
\bar{\nabla}_{[\sigma}C^{(n+1)}_{\alpha]\beta}-H^2\bar{g}_{\lambda[\sigma}D^{(n+1)\lambda}_{\phantom{(n+1)\lambda}\alpha]\beta}=0.
\end{array}
\ea
These equations admit a solution of the form
\ba
\begin{array}{l}
D^{(n+1)\lambda}_{\phantom{(n+1)\lambda}\alpha\beta}=2 \delta^\lambda_{(\alpha}\bar{\nabla}_{\beta)}f^{(n+1)} \, ,\\
C^{(n+1)}_{\alpha\beta}= \bar{\nabla}_\alpha \bar{\nabla}_\beta f^{(n+1)}\, ,
\end{array}
\ea
for a scalar function $f^{(n+1)}$ that is order $n+1$ in $h_{\mu\nu}$.  Once again, these terms are precisely of the form that can be absorbed by a redefinition of the gauge parameter \eqref{transf}, with $f =1+f^{(n+1)}$.  We thus conclude that
\be
C^{(n+1)}_{\alpha\beta}=0 \, , ~~~~~~  D^{(n+1)\lambda}_{\phantom{(n+1)\lambda}\alpha\beta}=0 \, .
\ee

Having established this, we now return to the $n$th-order closure condition and consider $B^{(n+1)\mu\nu}_{\phantom{(n+1)\mu\nu}\alpha\beta}$:
\begin{equation}
(\delta^{(0)}_{\phi}B^{(n+1)\mu\nu}_{\phantom{(n+1)\mu\nu}\alpha\beta})(\bar{\nabla}_{\mu}\bar{\nabla}_{\nu}\psi+H^2\bar{g}_{\mu\nu}\psi)-(\phi\leftrightarrow\psi)=(\bar{\nabla}_{\alpha}\bar{\nabla}_{\beta}+H^2\bar{g}_{\alpha\beta})\chi^{(n)} \, .
\end{equation}
Equivalently, we can write
\begin{multline}
\left(\frac{\partial B^{(n+1)\mu\nu}_{\phantom{(n+1)\mu\nu}\alpha\beta}}{\partial h_{\kappa\rho}}-\frac{\partial B^{(n+1)\kappa\rho}_{\phantom{(n+1)\kappa\rho}\alpha\beta}}{\partial h_{\mu\nu}}\right)(\bar{\nabla}_{\mu}\bar{\nabla}_{\nu}\psi+H^2\bar{g}_{\mu\nu}\psi)(\bar{\nabla}_{\kappa}\bar{\nabla}_{\rho}\phi+H^2\bar{g}_{\kappa\rho}\phi)\\
=(\bar{\nabla}_{\alpha}\bar{\nabla}_{\beta}+H^2\bar{g}_{\alpha\beta})\chi^{(n)}.
\end{multline}
As before, we operate on this equation with $\bar{\nabla}_{\sigma}$ and antisymmetrize over $\sigma$ and $\alpha$, with the result
\ba
\begin{array}{l}
\bar{\nabla}_{[\sigma}\left(\frac{\partial B^{(n+1)\mu\nu}_{\phantom{(n+1)\mu\nu}\alpha]\beta}}{\partial h_{\kappa\rho}}-\frac{\partial B^{(n+1)\kappa\rho}_{\phantom{(n+1)\kappa\rho}\alpha]\beta}}{\partial h_{\mu\nu}}\right)(\bar{\nabla}_{\mu}\bar{\nabla}_{\nu}\psi+H^2\bar{g}_{\mu\nu}\psi)(\bar{\nabla}_{\kappa}\bar{\nabla}_{\rho}\phi+H^2\bar{g}_{\kappa\rho}\phi)\\
+\left(\frac{\partial B^{(n+1)\mu\nu}_{\phantom{(n+1)\mu\nu}\beta[\alpha}}{\partial h_{\kappa\rho}}-\frac{\partial B^{(n+1)\kappa\rho}_{\phantom{(n+1)\kappa\rho}\beta[\alpha}}{\partial h_{\mu\nu}}\right)\Big(\bar{\nabla}_{\sigma]}(\bar{\nabla}_{\mu}\bar{\nabla}_{\nu}\psi+H^2\bar{g}_{\mu\nu}\psi)\,(\bar{\nabla}_{\kappa}\bar{\nabla}_{\rho}\phi+H^2\bar{g}_{\kappa\rho}\phi)\\
+\bar{\nabla}_{\sigma]}(\bar{\nabla}_{\kappa}\bar{\nabla}_{\rho}\phi+H^2\bar{g}_{\kappa\rho}\phi)\,(\bar{\nabla}_{\mu}\bar{\nabla}_{\nu}\psi+H^2\bar{g}_{\mu\nu}\psi)\Big)
=0\, .
\end{array}
\ea
This can hold for arbitrary $\phi$ and $\psi$ only if
\begin{equation} \label{eq:btensor_cond_n}
\frac{\partial B^{(n+1)\mu\nu}_{\phantom{(n+1)\mu\nu}\alpha\beta}}{\partial h_{\kappa\rho}}-\frac{\partial B^{(n+1)\kappa\rho}_{\phantom{(n+1)\kappa\rho}\alpha\beta}}{\partial h_{\mu\nu}}=0.
\end{equation}
This is the $n$th order part of (\ref{eq:btensor_cond}), the condition under which the higher-order terms in the tensor $B$ can be eliminated by means of a field redefinition. Eq.\ (\ref{eq:btensor_cond_n}) then implies that there exists a field redefinition that sets $B^{(n+1)\mu\nu}_{\phantom{(n+1)\mu\nu}\alpha\beta}=0$. We have thus demonstrated that the nontrivial $(n+1)$th order part of the PM gauge transformation vanishes under the assumptions \eqref{as1} and \eqref{as2}:
\begin{equation}
B^{(n+1)\mu\nu}_{\phantom{(n+1)\mu\nu}\alpha\beta}=0,\qquad D^{(n+1)\lambda}_{\phantom{(n+1)\lambda}\mu\nu}=0,\qquad C^{(n+1)}_{\mu\nu}=0.
\end{equation}

\bigskip

\subsection{Final results} \label{subsec:results}

For the fourth and higher order parts of the closure condition (\ref{eq:closure_cond_full}) we can apply the theorem of the previous section recursively, thereby concluding that all the higher order $B$, $D$ and $C$ tensors must vanish.  The full $B$, $D$ and $C$ tensors are
\ba
\begin{array}{lcl}
B^{\mu\nu}_{\phantom{\mu\nu}\alpha\beta}&=&\delta^{\mu}_{(\alpha}\delta^{\nu}_{\beta)},\\
D^{\lambda}_{\phantom{\lambda}\mu\nu}&=&0,\\
C_{\alpha\beta}&=&H^2\bar{g}_{\alpha\beta}+\gamma\,\bar{g}_{\alpha\beta}\left[F^{\lambda\mu\nu}F_{\lambda\mu\nu}-\frac{2}{(D-1)}F^{\lambda}F_{\lambda}\right] \, ,
\end{array}
\ea
with free parameter $\gamma$.  The unique candidate infinitesimal nonlinear PM gauge symmetry is then
\begin{equation} \label{eq:pm_candidate_symmetry}
\delta_{\phi} h_{\alpha\beta}=\left(\bar{\nabla}_{\alpha}\bar{\nabla}_{\beta}+H^2\bar{g}_{\alpha\beta}\right)\phi+\gamma\,\bar{g}_{\alpha\beta}\left[F^{\lambda\mu\nu}F_{\lambda\mu\nu}-\frac{2}{(D-1)}F^{\lambda}F_{\lambda}\right]\phi.
\end{equation}
%Note that it is not consistent to redefine the gauge function in the second term by absorbing the $h_{\mu\nu}$ dependence. Indeed, although both transformations $\delta_{\phi} h_{\alpha\beta}=\left(\bar{\nabla}_{\alpha}\bar{\nabla}_{\beta}+H^2\bar{g}_{\alpha\beta}\right)\phi$ and $\delta_{\phi} h_{\alpha\beta}=\bar{g}_{\mu\nu}\phi$ satisfy (trivially) the closure condition, the second one is not a symmetry of the free PM action.

An action that is separately invariant under the linear PM symmetry and a conformal-like transformation of the form $\delta_{\psi}^c h_{\alpha\beta}= \bar{g}_{\alpha\beta} \psi$ would trivially be invariant under this symmetry.  We note that the combination
\begin{equation}
F^{\lambda\mu\nu}F_{\lambda\mu\nu}-\frac{2}{(D-1)}F^{\lambda}F_{\lambda} \, ,
\end{equation}
in addition to being a PM invariant, is also invariant under the conformal-like transformations and is thus itself invariant under the transformation \eqref{eq:pm_candidate_symmetry}.  However, such a term is itself not a viable Lagrangian, since it doesn't have the ghost-free form of the PM theory \eqref{SPM}.  We emphasize that the nonlinear symmetry is not a trivial extension of the lowest order PM symmetry, in the sense that it is not obeyed by the free PM theory \eqref{SPM}.

\section{The action} \label{sec:action}
What sort of consistent action can realize such a symmetry?  The existence of a scalar gauge symmetry of the form
\begin{equation}
\delta h_{\mu\nu}=\hat{P}_{\mu\nu}\phi,
\end{equation}
where $\hat{P}_{\mu\nu}$ is an operator constructed locally from $h_{\mu\nu}$, implies that the equation of motion (EOM) $\mathcal{E}^{\mu\nu}\equiv \delta S/\delta h_{\mu\nu}$ must satisfy a corresponding Bianchi identity:
\begin{equation} \label{eq:bianchi_identity}
\hat{O}_{\mu\nu}\mathcal{E}^{\mu\nu}=0,
\end{equation}
where the operator $\hat{O}_{\mu\nu}$ is obtained from $\hat{P}_{\mu\nu}$ (and vice versa) by integration by parts.  Let us examine the full Bianchi identity that follows from (\ref{eq:pm_candidate_symmetry}):
\begin{equation} \label{eq:bianchi_full2}
\bar{\nabla}_{\mu}\bar{\nabla}_{\nu}\left(B^{\mu\nu}_{\phantom{\mu\nu}\alpha\beta}\mathcal{E}^{\alpha\beta}\right)-\bar{\nabla}_{\lambda}\left(B^{\mu\nu}_{\phantom{\mu\nu}\alpha\beta}D^{\lambda}_{\phantom{\lambda}\mu\nu}\mathcal{E}^{\alpha\beta}\right)+B^{\mu\nu}_{\phantom{\mu\nu}\alpha\beta}C_{\mu\nu}\mathcal{E}^{\alpha\beta}=0,
\end{equation}
Let us write the EOM as a linear plus nonlinear parts, $\mathcal{E}^{\alpha\beta}=\mathcal{E}^{(1)\alpha\beta}+\Delta\mathcal{E}^{\alpha\beta}$, where $\left(\bar{\nabla}_{\alpha}\bar{\nabla}_{\beta}+H^2\bar{g}_{\alpha\beta}\right)\mathcal{E}^{(1)\alpha\beta}=0$.  Eq.\ (\ref{eq:bianchi_full2}) yields
\begin{equation}
\left(\bar{\nabla}_{\alpha}\bar{\nabla}_{\beta}+H^2\bar{g}_{\alpha\beta}\right)\Delta\mathcal{E}^{\alpha\beta}+\gamma\,\widetilde{C}^{(2)}\,\bar{g}_{\alpha\beta}\,\mathcal{E}^{(1)\alpha\beta}+\gamma\,\widetilde{C}^{(2)}\,\bar{g}_{\alpha\beta}\,\Delta\mathcal{E}^{\alpha\beta}=0 \, ,
\end{equation}
where $\widetilde{C}^{(2)} = F^{\lambda\mu\nu}F_{\lambda\mu\nu}-\frac{2}{(D-1)}F^{\lambda}F_{\lambda}$. Considering this expression perturbatively, we observe,
\begin{equation}
\begin{split}
\left(\bar{\nabla}_{\alpha}\bar{\nabla}_{\beta}+H^2\bar{g}_{\alpha\beta}\right)\mathcal{E}^{(2)\alpha\beta}&=0\, ,\\
\left(\bar{\nabla}_{\alpha}\bar{\nabla}_{\beta}+H^2\bar{g}_{\alpha\beta}\right)\mathcal{E}^{(k+2)\alpha\beta}&=-\gamma\,\widetilde{C}^{(2)}\,\bar{g}_{\alpha\beta}\,\mathcal{E}^{(k)\alpha\beta} \, ,
\end{split}
\end{equation}
for $k \geq 1$.  Note that $\gamma$ plays the role of a dimensionless coupling constant, as the terms with higher powers of $h_{\mu\nu}$ are proportional to higher powers of $\gamma$.

Consider the Bianchi identity that constrains the cubic EOM.  If we take the lowest order Lagrangian to be the free, ghost-free PM theory \eqref{eq:tensorF}, the Bianchi identity reads explicitly
\begin{equation} 
\label{eq:bianchi_cubic_eom}
\left(\bar{\nabla}_{\alpha}\bar{\nabla}_{\beta}+H^2\bar{g}_{\alpha\beta}\right)\mathcal{E}^{(3)\alpha\beta}=\gamma\,(D-2)\bar{\nabla}_{\sigma}F^{\sigma}\left[F^{\lambda\mu\nu}F_{\lambda\mu\nu}-\frac{2}{(D-1)}F^{\lambda}F_{\lambda}\right]\,.
\end{equation}
Conceivably, a two derivative action that is quartic in the fields ${\cal L}^{(4)}$ might be able to satisfy such an equation.

In order to check, we perform a brute force, perturbative analysis.  In fact, our analysis is more general than that required by \eqref{eq:bianchi_cubic_eom}.  We take as a starting point the quadratic PM action \eqref{eq:tensorF}.   We then consider every possible cubic and quartic interaction with at most two derivatives.  In addition, we consider the most generic linear and quadratic extensions of the gauge transformation that themselves have at most two derivatives.  We then determine whether a choice of coefficients exists so that eq \eqref{eq:bianchi_identity} can be satisfied order by order.   We find that no such action exists.  Since the cubic case was already considered in \cite{Zinoviev:2006im} and the quartic result was already stated there as well, we only briefly summarize our findings here:

The cubic Lagrangian $\mathcal{L}^{(3)}$ is the unique two-derivative term that satisfies the second order identity,
\begin{equation} 
\label{eq:bianchi_order2}
\hat{O}^{(0)}_{\mu\nu}\frac{\delta\mathcal{L}^{(3)}}{\delta h_{\mu\nu}}+\hat{O}^{(1)}_{\mu\nu}\frac{\delta\mathcal{L}^{(2)}}{\delta h_{\mu\nu}}=0,
\end{equation}
Here $\hat{O}^{(0)}_{\mu\nu}$ is given by the lowest order PM symmetry, and $\mathcal{L}^{(2)}$ is the free PM Lagrangian.   Allowing for non-canonical derivative interactions, we find only one cubic action $\mathcal{L}^{(3)}$ exists that satisfies this expression and only when $D=4$, consistent with the results of \cite{Zinoviev:2006im,deRham:2013wv}.

The part of the Bianchi identity containing three powers of $h_{\mu\nu}$ is given by
\begin{equation} \label{eq:bianchi_order3}
\hat{O}^{(0)}_{\mu\nu}\frac{\delta\mathcal{L}^{(4)}}{\delta h_{\mu\nu}}+\hat{O}^{(1)}_{\mu\nu}\frac{\delta\mathcal{L}^{(3)}}{\delta h_{\mu\nu}}+\hat{O}^{(2)}_{\mu\nu}\frac{\delta\mathcal{L}^{(2)}}{\delta h_{\mu\nu}}=0.
\end{equation}
The generic quartic Lagrangian contains 5 zero-derivative contractions with four powers of $h_{\mu\nu}$. We choose to write the two-derivative terms in contractions of the form $hh\bar{\nabla}h\bar{\nabla}h$. There are 43 such contractions; however, five of them can be show to be redundant via identities.  Thus the generic form of $\mathcal{L}^{(4)}$ contains a total of 43 free parameters. For the operator $\hat{O}^{(2)}_{\mu\nu}$ we find 4 terms with no derivatives plus 68 with two derivatives, for a total of 72 parameters to be determined. The Bianchi identity (\ref{eq:bianchi_order3}) then contains contractions with zero, two and four derivatives with three powers of $h_{\mu\nu}$. We count 16 contractions of the form $hh\bar{\nabla}\bar{\nabla}\bar{\nabla}\bar{\nabla}h$, 50 contractions of the form $h\bar{\nabla}h\bar{\nabla}\bar{\nabla}\bar{\nabla}h$, 45 contractions of the form $h\bar{\nabla}\bar{\nabla}h\bar{\nabla}\bar{\nabla}h$, 65 contractions of the form $\bar{\nabla}h\bar{\nabla}h\bar{\nabla}\bar{\nabla}h$, 12 contractions of the form $hh\bar{\nabla}\bar{\nabla}h$, 16 contractions of the form $h\bar{\nabla}h\bar{\nabla}h$, and 3 contractions of the form $hhh$. The total number of constraints is therefore 207, which involve 115 
parameters (116 in $D=4$).

We then find that no set of nonzero coefficients exists that solves the constraints, except for the trivial ones that arise from field redefinitions of the free PM Lagrangian. In particular, the cubic Lagrangian $\mathcal{L}^{(3)}$ inevitably generates an obstruction at the next order in the Bianchi identity.  Furthermore, even if cubic interactions are absent, there exist no quartic interactions with up to two derivatives that exhibit a gauge symmetry.

Note that, due to the recursive relation \eqref{eq:bianchi_cubic_eom} the absence of a quartic Lagrangian means that no nonlinear two-derivative Lagrangian can realize the nonlinear PM symmetry with $\gamma \neq 0$.  In other words, one could conceive of action that is two derivatives in the fields, has no cubic or quartic terms, and yet somehow realizes a nonlinear symmetry via higher order terms.  Eq \eqref{eq:bianchi_cubic_eom} rules out this case: this equation cannot be satisfied if $\mathcal{E}^{(3)\alpha\beta}=0$.

\bigskip

\section{Discussion} \label{sec:discussion}

The closure condition \eqref{eq:closure_cond_full} places powerful constraints on any nonlinear extension of the partially massless symmetry, while allowing one to remain entirely agnostic about the form of the invariant Lagrangian.  The basic assumption of this paper was that the partially massless symmetry itself contains no more than two derivatives of the fields.  With this assumption and using the closure condition we were able to identify a unique nonlinear partially massless symmetry.  We could then show that no consistent Lagrangian which contains at most two derivatives can realize this symmetry.  

For the closure condition, we have demanded that two gauge symmetries close to another gauge symmetry.  More generally, it's potentially consistent for the gauge symmetries to close to a gauge symmetry plus an on-shell trivial symmetry.   While this is not the situation for the gauge symmetries of massless spin-1 and spin-2 fields, this occurs, for example, in the case of supersymmetry without auxiliary fields.\footnote{We are grateful to Kurt Hinterbichler for pointing this out.}  We have checked to see if such considerations modify our results.  At lowest order, generalizing the closure condition to allow for trivial on-shell symmetries leads only to new symmetry terms that are themselves on-shell trivial.  However, it remains possible that this generalization could lead to a wider family of candidate symmetries at higher order.

%In this paper we have investigated the integrability condition for the infinitesimal PM gauge symmetry---namely that the commutator of two gauge transformations must be a gauge transformation---with the goal of classifying the possible symmetries of a putative nonlinear theory of PM gravity. The general solution to this condition, shown in eq.\ (\ref{eq:pm_candidate_symmetry}), consists of a single term, quadratic in $h_{\mu\nu}$, beyond the lowest-order symmetry of the free PM action. The Bianchi identity that follows from the full symmetry gives a set of recurrence relations that constrain the equations of motion at each order in the field, and imply that the higher order nonlinear terms are proportional to higher powers of a ``coupling constant'' $\gamma_1$. We have also shown, in particular, that there exist no quartic Lagrangians with up to two derivatives satisfying this constraint when $\gamma_1\neq0$. Morover, even without making any assumptions regarding the forms of the Lagrangian and the gauge transformation (except for the condition that they contain no terms with more than two derivatives), we found that a Bianchi identity does not exist at the quartic order in the interactions, thus generalizing the no-go results of \cite{deRham:2013wv} to include noncanonical (i.e.\ non-GR) kinetic terms.

The candidate nonlinear symmetry (\ref{eq:pm_candidate_symmetry}) has some curious properties that distinguishes it from its GR and Yang-Mills counterparts. The symmetry not only has the feature of being Abelian, $[\delta_{\phi},\delta_{\psi}]h_{\alpha\beta}=0$, but it is also {\it nilpotent},
\begin{equation}
\delta_{\phi}\delta_{\psi}h_{\alpha\beta}=0.
\end{equation}
This means that the transformation solves the closure condition in a rather trivial way despite it being nonlinear. A consequence of the nilpotency property is that the infinitesimal transformation can be trivially integrated to yield the corresponding {\it finite} gauge transformation.
%\begin{equation}
%h_{\alpha\beta}\to h_{\alpha\beta}'=h_{\alpha\beta}+(\bar{\nabla}_{\alpha}\bar{\nabla}_{\beta}+H^2\bar{g}_{\alpha\beta})\Phi+\gamma\,\bar{g}_{\alpha\beta}\left[F^{\lambda\mu\nu}F_{\lambda\mu\nu}-\frac{2}{(D-1)}F^{\lambda}F_{\lambda}\right]\Phi,
%\end{equation}
%where now the gauge function $\Phi$ need not be infinitesimal. 
Despite these simple properties, we emphasize that the nonlinear symmetry is not a trivial extension of the lowest order PM symmetry, in the sense that it is not obeyed by the free PM theory.

Another interesting (if suspicious) feature of the nonlinear symmetry is that it persists in the flat space limit.  In the limit that $H \rightarrow 0$, the lowest order part of the symmetry, i.e., $\delta h_{\mu\nu} = \partial_\mu\partial_\nu \phi$ is simply a diffeomorphism with gauge parameter $\xi_\mu = \partial_\mu \phi$.  Thus this gives us nothing unexpected.  However, the nonlinear part of the PM symmetry is nonzero in the flat space limit and yet cannot be written as a diffeomorphism.  This would seem to imply the existence of some sort of new scalar gauge symmetry for a spin-2 field on a flat background.

If this symmetry can at all be realized by a Lagrangian, either higher derivative terms or additional fields are required.  The method we have used here gives only the form of the symmetry and tells us little else about other properties of the invariant Lagrangian, such its health or stability.  Thus even if a higher derivative action possesses the nonlinear PM symmetry, recent results \cite{deRham:2013tfa} cast doubt on whether such an action can be ghost-free.

Finally, we note that imposing the closure condition on the combination of the PM symmetry plus diffeomorphisms, rather than on the PM symmetry alone can potentially allow for more general symmetries than those found here.  Such a condition would be appropriate for a partially massless particle coupled to gravity.  Such a possibility was considered in \cite{Joung:2014aba} and a no-go result was obtained.  Our approach is more general than the one of \cite{Joung:2014aba} in that we do not assume any specific form for the nonlinear Lagrangian.  It's possible that the application of the approach used here might lead to a nonlinear symmetry in this case, though it will not necessarily lead to a Lagrangian.  We leave this for future work.

%Finally, we can consider the consequence of relaxing the assumption of having only two derivatives in the nonlinear extension of the partially massless symmetry.  Generically, a theory obeying this type of symmetry would contain an ever increasing number of derivatives in the higher-order interactions.  However, it is not inconceivable that the full action may have a resummed, non-polynomial form.

%Another question concerns the existence of nonlinear higher-derivative theories of PM gravity. We remarked in section \ref{sec:pm_gravity} that one may look for consistent actions depending only on the invariant tensor $F_{\lambda\mu\nu}$, for which the usual symmetry of free PM gravity is exact. Alternatively, we also noted that it is an open question whether the Bianchi identity can be satisfied by some higher-derivative theory in the nontrivial case $\gamma_1\neq0$. Finally, if one is willing to lift all restrictions regarding the number of derivatives in the action, it would be only natural to do the same at the level of the gauge transformation. A theory of this type would be rather exotic, as the Bianchi identity corresponding to a symmetry with more than two derivatives would yield---at least generically---an ever increasing number of derivatives in the higher-order interactions, although it is not inconceivable that the full action may have a resummed, non-polynomial form.

\acknowledgments
We would like to thank Kurt Hinterbichler, Alberto Nicolis, Riccardo Rattazzi, Andrew Tolley and Bob Wald for many productive conversations.  SGS and RAR are supported by DOE grant DE-SC0011941.

\bibliographystyle{JHEP}
\bibliography{nonlinearPM}

\end{document}